\documentclass[usenatbib]{mnras}
\usepackage[english]{babel}

\pdfoutput=1
\usepackage{amsfonts}
\usepackage{amsmath}
\usepackage{amssymb}
\usepackage{xcolor}
\usepackage{enumitem}
\usepackage{graphicx}
\usepackage{gensymb}
\usepackage{fancyhdr}
\usepackage{times}
\usepackage{widetext}
\usepackage{ulem}
\usepackage{caption}
\usepackage{subcaption} 
\usepackage{float}

\newcommand{\LCDM}{$\Lambda$CDM}

\graphicspath{{./Figures/}}

\begin{document}

\title[Non-linear matter power spectra for arbitrary cosmologies]{On the road to percent accuracy II: calibration of the non-linear matter power spectrum for arbitrary cosmologies}

\author[Giblin et al.]
{Benjamin Giblin$^1$\thanks{bengib@roe.ac.uk}, Matteo Cataneo$^1$, Ben Moews$^1$ and Catherine Heymans$^1$   \\
$^1$ Scottish Universities Physics Alliance, Institute for Astronomy, University of Edinburgh, Blackford Hill, Edinburgh, EH9 3HJ, UK \\ }
\maketitle

\begin{abstract}
We introduce an emulator approach to predict the non-linear matter power spectrum for broad classes of beyond-$\Lambda$CDM cosmologies, using only a suite of $\Lambda$CDM $N$-body simulations. By including a range of suitably modified initial conditions in the simulations, and rescaling the resulting emulator predictions with analytical `halo model reactions', accurate non-linear matter power spectra for general extensions to the standard $\Lambda$CDM model can be calculated. We optimise the emulator design by substituting the simulation suite with non-linear predictions from the standard {\sc halofit} tool.  We review the performance of the emulator for artificially generated departures from the standard cosmology as well as for theoretically motivated models, such as $f(R)$ gravity and massive neutrinos. For the majority of cosmologies we have tested, the emulator can reproduce the matter power spectrum with errors $\lesssim 1\%$ deep into the highly non-linear regime.  This work demonstrates that with a well-designed suite of $\Lambda$CDM simulations, extensions to the standard cosmological model can be tested in the non-linear regime without any reliance on expensive beyond-$\Lambda$CDM simulations.
\end{abstract} 
\begin{keywords}
   Cosmology: theory -- Cosmology: cosmological parameters -- Surveys -- Methods: statistical
\end{keywords}

\section{Introduction}\label{sec:introduction}

Large cosmological data sets have ushered in the era of ``precision cosmology", in which parameters describing the \LCDM{} model are known with uncertainties of only a few percent. This increased precision has recently brought to light some level of discordance between early- and late-time cosmological probes. In particular, all three major weak lensing surveys \citep{Hildebrandt_etal_2016,Hikage_etal_2018, DES_Cosmic_Shear_Yr1} infer lower values of $\sigma_8 \sqrt{\Omega_{\rm{m}}}$ compared to those obtained from observations of the temperature and polarisation of the cosmic microwave background \citep[CMB;][]{Planck_2018}. Significant tensions have also arisen between the value of the Hubble constant derived from the CMB and its direct local measurement via the distance-redshift relation calibrated with Cepheid variables and Type Ia supernovae \citep{Riess_etal_2016,Riess_etal_2018}. If not the product of an unlikely statistical fluctuation, then these discrepancies may be ascribed either to unaccounted systematic uncertainties in the analyses or to an incompleteness of the concordance cosmology \citep[see, e.g.,][]{Verde_etal_2013,Joudaki_etal_2016, Mortsell_Dhawan_2018,DES_etal_2018}.

In the event that the data prefer a revision of the standard cosmological model, accurate predictions of large-scale structure statistics for alternative cosmologies to $\Lambda$CDM become very valuable. However, we currently lack a method to predict these observables for general extensions to \LCDM{} and over the range of scales relevant for ongoing and future galaxy surveys \citep{Euclid_etal_2011,LSST_etal_2012, Hildebrandt_etal_2016, DES_Clustering_Cosmic_Shear_Yr1, Taylor_etal_2018}. This means we cannot take advantage of the wealth of information contained in the non-linear scales without resorting to computationally expensive numerical simulations.  

It is however infeasible to simulate all possible combinations of cosmological parameters, even within the \LCDM{} paradigm. For current applications, fitting functions calibrated on a small number of $N$-body simulations are good enough to obtain unbiased constraints \citep{Smith_etal_2003,Takahashi_etal_2012,Mead_etal_2015}. On the other hand, future all-sky surveys will place more stringent requirements on our theoretical predictions, which makes emulator techniques a preferable strategy as far as accuracy is concerned. The basic idea is to interpolate between the output of $\sim$10--100 $N$-body simulations of cosmologies associated with carefully-chosen input parameters \citep{Habib_etal_2007,Schneider_etal_2008, Heitmann_etal_2009, Heitmann_etal_2016, Lawrence_etal_2017, EuclidEmulator_etal_2018, Winther_etal_2019, Harnois-Deraps_etal_2019}. Previous implementations of this methodology have been applied either to flat, massless neutrinos (i.e. vanilla) \LCDM{}, or to a few specific extensions, that is models with massive neutrinos, evolving dark energy with a linearly parameterised equation of state or $f(R)$ gravity. The computational expense of simulating modified gravity theories, however, can be orders of magnitude larger than that of the standard model \citep{Llinares_2018}. Consequently, there currently exists no emulator capable of making non-linear predictions accurate at the level of $\sim$1\% across a broad range of beyond-\LCDM{} cosmologies. This severely limits our capacity for constraining these alternative models with data probing the growth of structure on cosmic scales.

In this paper we complement the work of~\citet[][referred to as `C19' henceforth]{Cataneo_etal_2019}, who formulated an efficient method to compute the non-linear matter power spectrum in a broad range of extensions to the standard cosmology with the desired percent accuracy. The C19 strategy consists of rescaling the non-linear power spectrum of a \textit{tailored} \LCDM{}-{\it like} model -- the {\it pseudo} cosmology -- by an analytical function encapsulating the non-linear physics beyond the vanilla cosmology -- the {\it halo model reaction}. By construction the pseudo cosmology shares the linear power spectrum of the {\it real} beyond-\LCDM{} cosmology. 

For \LCDM{} extensions exhibiting a scale-independent linear growth of structure, the corresponding pseudo non-linear power spectrum can be obtained from a \LCDM{} emulator by matching the amplitude of mass fluctuations, $\sigma_8$, to that of the real cosmology at the redshift of interest. In general, however, structures in alternative cosmological scenarios grow at different rates on different scales, which makes traditional emulators unfit for this task. Here, we present a new Gaussian process emulator designed to compute the pseudo non-linear matter power spectrum of arbitrary cosmologies, including both scale-independent and scale-dependent linear growth. We show that a suite of $\sim$1000 \LCDM{}-{\it like} simulations is enough to predict with percent accuracy the pseudo non-linear power spectrum of popular cosmologies which have scale-dependent linear growth, such as $f(R)$ gravity and massive neutrino models. We also obtain similar performance for synthetic cosmological models created by rescaling the linear \LCDM{} power spectra by smooth arbitrary functions. 

The combination of the C19 halo model reactions with the emulator developed in this work will hence provide accurate predictions of the non-linear matter power spectrum in a broad class of cosmological models, including modified gravity, dark energy and massive neutrinos, for use in cosmic shear analyses \citep[see, e.g.,][]{DES_Cosmic_Shear_Yr1, Hikage_etal_2018, Hildebrandt_etal_2018}. The fact that galaxies are biased tracers of the underlying matter distribution means that additional emulator complexity, including parameters to describe the halo occupation distribution, would be required for this methodology to be used in future clustering analyses \citep[see, e.g.,][]{Zhai_etal_2019}. There is also the potential to use this technique to model the impact of baryonic feedback.

This paper is organised as follows. In Section \ref{sec:Beyond_LCDM} we examine the expected range of deviations in the linear matter power spectrum of viable extensions to the \LCDM{} cosmology, and outline the C19 reaction framework for modelling the non-linear power spectrum. In Section \ref{sec:method} we detail the emulator methodology for arbitrary input linear power spectra, where we define `arbitrary' to mean any power spectra which deviates smoothly from the standard model approximately within the range allowed by observational constraints. We also describe the construction of our suite of surrogate numerical simulations used as a training set for the emulator, which we design to encompass a reasonable range of $\Lambda$CDM extensions. In Section \ref{sec:Results} we present the performance of our emulator and determine the minimum number of simulations we need to reach the accuracy required by the high-quality data from the next generation of all-sky imaging surveys. Finally, we conclude and discuss the impact of our findings in Section \ref{sec:Conclusions}.

\section{Beyond vanilla $\Lambda$CDM} \label{sec:Beyond_LCDM}

The late-time phenomenology of the concordance model -- a globally flat Universe with a matter-energy content dominated by cold dark matter, baryonic matter and the cosmological constant -- is entirely described by five \textit{base} parameters, $\boldsymbol{\pi}_{\Lambda} \equiv \{ \Omega_{\rm{b}}h^2,\Omega_{\rm{m}}h^2, h, n_{\rm s}, \ln(10^{10}A_{\rm s}) \}$, where $\Omega_{\rm{b}}$ is the current fraction of energy-density in baryonic matter, $\Omega_{\rm{m}}$ is the present-day total matter energy-density fraction, $h$ is the dimensionless Hubble constant, and $n_{\rm s}$ and $A_{\rm s}$ are, respectively, the slope and amplitude of the primordial scalar power spectrum. Under the assumption of the standard cosmology, measurements of the CMB constrain the radiation content (i.e. photons and three massless neutrino species) to better than one part in a thousand \citep{Fixsen_2009}.

Changes to the dark sector and to the law of gravity have profound implications for the formation of structures in the Universe over a wide range of scales. Departures from the standard cosmology, therefore, leave distinctive features on the statistical quantities measuring the clustering of matter, such as the two-point correlation function, or its Fourier transform, the power spectrum $P(k)$. The landscape of \LCDM{} extensions is exceptionally vast, and in this work we only consider a family of $f(R)$ gravity theories \citep{Hu_Sawicki_2007} and massive neutrino cosmologies~\citep{Lesgourgues_Pastor_2006}, both with a \LCDM{} background. However, with little or no modification our methodology (see Section~\ref{sec:method}) can be applied to a much broader class of models, including non-standard dark matter candidates ~\protect\citep{Marsh_Silk_2014,Schneider_2015,Cyr-Racine_etal_2016,Marsh_2016,Poulin_etal_2016,Hlovzek_etal_2017,Dakin_etal_2019,Thomas_etal_2019}, extra relativistic degrees of freedom \citep{Baumann_etal_2018}, and Horndeski theories \citep{Zumalac_etal_2017}.

What makes models like $f(R)$ gravity and massive neutrino cosmologies interesting is their scale-dependent linear growth of structure. More specifically, $f(R)$ theories enhance the growth on scales comparable to, or smaller than, the Compton wavelength, which at $z=0$ reads \citep{Hu_Sawicki_2007}
\begin{equation}
\lambda_{{\rm C}0} \approx 30 \sqrt{\frac{(n+1)}{4-3\Omega_{\rm m}}\frac{|f_{R0}|}{10^{-4}}} \quad h^{-1}\rm{Mpc},
\end{equation}
where $f_{R0}$ defines the extent of the departure from General Relativity, recovered for $f_{R0} = 0$. We set $n=1$ and use $|f_{R0}|=10^{-4},10^{-5},10^{-6}$ (referred to as F4, F5 and F6 in the following), which brackets the range of values giving rise to interesting cosmological behaviours, with F4 already strongly disfavoured by current data~\citep{Terukina_etal_2014, Lombriser_2014, Cataneo_et_al_2015, Liu_etal_2016, Alam_etal_2016}.

On the other hand, the high thermal velocities of massive neutrinos prevent them from clustering on scales smaller than the free-streaming length, $\lambda_{\rm FS}$. This results in the suppression of the growth of structure on scales smaller than the maximum free-streaming length, which is defined at the time of the non-relativistic transition (after recombination for $m_\nu \lesssim 0.5 \, {\rm eV}$), and is given by~\citep{Lesgourgues_Pastor_2006}
\begin{equation}
\lambda_{{\rm FS}}^{{\rm max}} \approx 350 \sqrt{\frac{1}{\Omega_{\rm m}}\frac{1 \, {\rm eV}}{m_\nu}} \quad h^{-1}\rm{Mpc}.
\end{equation}

For simplicity, we will consider cosmologies including two massless neutrinos and one massive neutrino with $m_\nu = 0.05, 0.1, 0.2, 0.4 \, {\rm eV}$, where the smallest value is close to the minimum sum of neutrino masses consistent with neutrino oscillation experiments \citep{Forero_etal_2014}, and the largest value is outside the 95\% confidence region found by~\cite{Planck_2018}. In addition, for all these extensions we keep the base cosmological parameters fixed to their \LCDM{} values, and compensate for the presence of massive neutrinos by reducing the cold dark matter density as $\Omega_{\rm c} \rightarrow \Omega_{\rm m} - \Omega_{\rm b} - \Omega_{\nu}$, where
\begin{equation}
\Omega_\nu h^2 = \frac{\sum m_\nu}{93.14 \, {\rm eV}}.
\end{equation}

\begin{figure}
\centering
\includegraphics[width=0.5\textwidth]{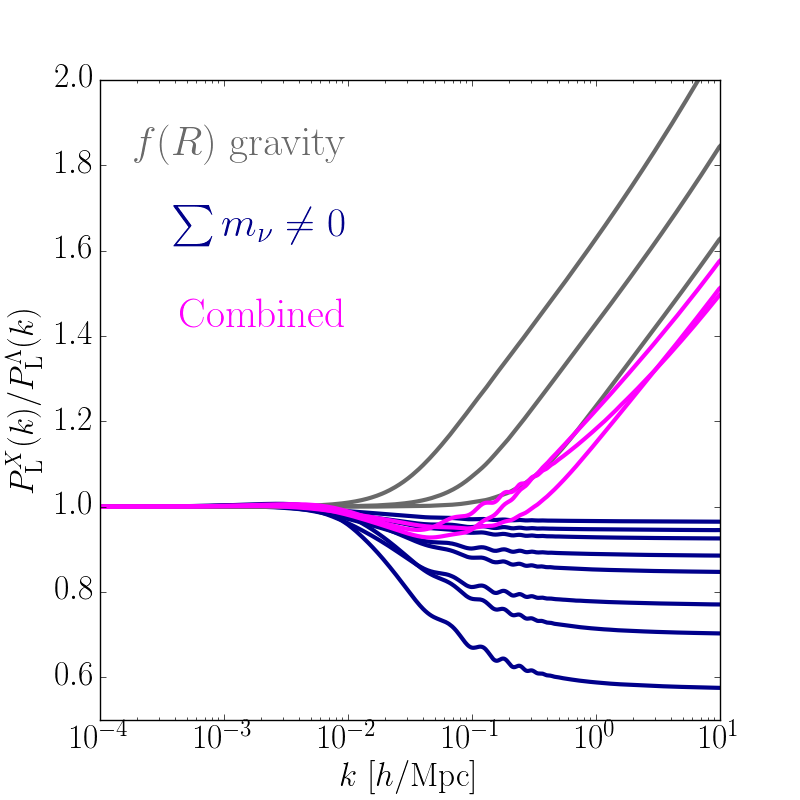} 
\caption{Beyond-\LCDM{} ($P_{\rm L}^X$) to \LCDM{} ($P_{\rm L}^\Lambda$) linear matter power spectrum ratios for the physical models in Table~\ref{tab:PhysMod_Table} at $z=0$. Pure $f(R)$ gravity (in grey) and massive neutrino cosmologies (in blue) alter the clustering of matter in similar but  opposite ways. This degeneracy is clearly visible for the hybrid models (in magenta), where both extensions are active. These curves will be used as guidelines to design our emulator.} \label{fig:PhysicalShapes}
\end{figure}

Table~\ref{tab:PhysMod_Table} lists all models, collectively referred to as \textit{physical models,} used to test our emulation scheme described in Section~\ref{sec:method}, including hybrid cosmologies with massive neutrinos and modified gravity. For these extensions, the linear power spectrum deviations from the standard predictions are obtained with {\sc mgcamb}\footnote{\url{http://aliojjati.github.io/MGCAMB/mgcamb.html}}\citep{Zhao_etal_2009,Hojjati_etal_2011}. Figure~\ref{fig:PhysicalShapes} shows a $z=0$ example of beyond-\LCDM{} to \LCDM{} linear matter power spectrum ratios. 

\begin{table}
\caption{The massive neutrino, $f(R)$ gravity and hybrid models used in this work to design and test our emulator scheme, each labelled according to the sum of neutrino masses ($\sum m_\nu$) and/or the strength of the deviation from standard gravity ($|f_{R0}|$).  We consider two base \LCDM{} cosmologies: (A) $\Omega_{\rm{b}}h^2=0.0225,\Omega_{\rm{m}}h^2=0.1382, h=0.6898, n_{\rm s}=0.969, \ln(10^{10}A_{\rm s})=3.195$; and (B) $\Omega_{\rm{b}}h^2=0.0173,\Omega_{\rm{m}}h^2=0.0864, h=0.6, n_{\rm s}=0.969, \ln(10^{10}A_{\rm s})=3.807$. Note that (B) sits several standard deviations away from the~\protect\cite{Planck_2018} best fit cosmology, thus working as stress test for our method.}
\begin{center}
\begin{tabular}{cccc}
\hline
\hline
model & $\sum m_\nu$ & $|f_{R0}|$ & base parameters \\
\hline 
MNU\_0.05 & $0.05 \, {\rm eV}$ & -- & (A) \\
MNU\_0.1 & $0.1 \, {\rm eV}$ & -- & (A) \\
MNU\_0.2 & $0.2 \, {\rm eV}$ & -- & (A) \\
MNU\_0.4 & $0.4 \, {\rm eV}$ & -- & (A) \\
xMNU\_0.05 & $0.05 \, {\rm eV}$ & -- & (B) \\
xMNU\_0.1 & $0.1 \, {\rm eV}$ & -- & (B) \\
xMNU\_0.2 & $0.2 \, {\rm eV}$ & -- & (B) \\
xMNU\_0.4 & $0.4 \, {\rm eV}$ & -- & (B) \\
F4 & -- & $10^{-4}$ & (A) \\
F5 & -- & $10^{-5}$ & (A) \\
F6 & -- & $10^{-6}$ & (A) \\
F4-MNU\_0.4 & $0.4 \, {\rm eV}$ & $10^{-4}$ & (A) \\
F5-MNU\_0.2 & $0.2 \, {\rm eV}$ & $10^{-5}$ & (A) \\
F6-MNU\_0.1 & $0.1 \, {\rm eV}$ & $10^{-6}$ & (A) \\
\hline
\hline
\end{tabular}
\end{center}
\label{tab:PhysMod_Table}
\end{table}%

\subsection{The halo model reaction framework} \label{subsec:C19}

By unlocking the information stored in the statistical distribution of matter on small scales we can substantially increase our sensitivity to physics beyond the vanilla cosmology, be it new particles/fluids, modifications to the theory of gravity, or astrophysics \citep[see, e.g.][]{Huterer_etal_2015,Copeland_etal_2018,Nori_Baldi_2018,Heymans_Zhao_2018,Schneider_etal_2019}. Not surprisingly, then, the non-linear matter power spectrum is found at the core of all cosmological analyses hinged on galaxy survey data ~\protect\citep{Parkinson_etal_2012,Kitching_etal_2014,Gil-Martin_etal_2016,Hildebrandt_etal_2016, vanUitert_etal_2017,Joudaki_etal_2017, DES_Clustering_Cosmic_Shear_Yr1, Gil-Martin_etal_2018,Hikage_etal_2018}. To take full advantage of the exquisite observations from forthcoming experiments \citep{Euclid_etal_2011, LSST_etal_2012, DESI_etal_2013}, theoretical predictions must reach percent level accuracy deep in the non-linear regime of structure formation~\citep{Taylor_etal_2018}. At present, however, no method is at the same time computationally efficient, accurate and flexible enough to be employed in future analyses aimed at constraining physics beyond the standard paradigm ~\protect\citep{Lesgourgues_etal_2009,Bird_etal_2012, Heitmann_etal_2014,Brax_Valagaes_2013,Zhao_2014, Blas_etal_2014,Massara_etal_2014,Levi_Vlah_2016, Lawrence_etal_2017,Aviles_Cervantes-Cota_2017,Senatore_Zaldarriaga_2017, Bose_etal_2018,EuclidEmulator_etal_2018, Heisenberg_Bartelmann_2019,Winther_etal_2019}.

The halo model reaction framework first proposed by C19, and based on the work of \citet{Mead_2017}, stands out as a promising, practical solution to the long-standing problem of predicting the non-linear matter power spectrum with the required accuracy and speed, while also being readily applicable to a variety of cosmological scenarios. Within this approach, the non-linear power spectrum of a given \textit{real} cosmology takes the form
\begin{equation}\label{eq:HMR_frame}
P_{\rm NL}^{\rm real}(k,z) = \mathcal{R}(k,z) \times P_{\rm NL}^{\rm pseudo}(k,z),
\end{equation}
where the \textit{pseudo} cosmology is defined as a \LCDM{} cosmology sharing the same linear matter power spectrum of the \textit{real} cosmology, that is
\begin{equation}
P_{\rm L}^{\rm pseudo}(k,z) = P_{\rm L}^{\rm real}(k,z).
\end{equation} 
The \textit{reaction},
\begin{equation} 
\mathcal{R}(k,z) \equiv \left. \frac{P_{\rm NL}^{\rm real}(k,z)}{P_{\rm NL}^{\rm pseudo}(k,z)} \right|_{\rm HM},
\end{equation} 
captures the non-linearities sourced by late-time physics beyond \LCDM{}, and is predicted with the halo model (HM) and perturbation theory~\citep[see][for details]{Cataneo_etal_2019}. In taking the ratio of the two halo model predictions,~\cite{Mead_2017} showed that the impact of the known inaccuracies in the individual power spectra is significantly reduced. On linear scales and for \textit{real} \LCDM{} cosmologies one has the trivial relation $\mathcal{R} = 1$. C19 found that the modelling approach summarised by equation \ref{eq:HMR_frame} is accurate at the percent level for a broad range of general extensions to the standard model.

Here, we formulate a method to provide predictions for $P_{\rm NL}^{\rm pseudo}$ that are more accurate than any semi-analytical prescription based on the halo model. It is this accurate estimate that is then used to calibrate the reaction, $\mathcal{R}$, in equation~\ref{eq:HMR_frame} in order to derive the non-linear matter power spectrum of our target beyond-\LCDM{} cosmology, $P_{\rm NL}^{\rm real}$. To this end, we develop a novel emulation scheme where in addition to the base \LCDM{} parameters we use the shape of the linear matter power spectrum as input. This work is intended as a first feasibility study, and as such we approximate the output of otherwise expensive \textit{pseudo} $N$-body simulations with the {\sc halofit} fitting functions~\citep{Takahashi_etal_2012} implemented in a modified version of the public Einstein-Boltzmann solver {\sc camb}\footnote{\url{https://github.com/cmbant/CAMB}} \citep{Lewis_etal_2000}. In practice, for a model $X$, {\sc camb} reads in both $\boldsymbol{\pi}_{\Lambda}$ and the externally generated $P_{\rm L}^X/P_{\rm L}^\Lambda$ ratio, evaluates $P_{\rm L}^{\rm pseudo}$ and gives it to {\sc halofit}, which finally provides the desired non-linear quantity. The grey curves in the upper panel of Figure~\ref{fig:Boostzdep} show the \textit{pseudo} non-linear power spectra for the physical models in Table~\ref{tab:PhysMod_Table}, where we used the set (A) of base \LCDM{} parameters also for the xMNU cosmologies. Note that despite {\sc halofit} predictions being approximate, this strategy is theoretically consistent, in that \textit{pseudo} cosmologies are simply \LCDM{} cosmologies with non-standard initial conditions. Our approach enables a rapid design, construction and performance assessment of the emulator~\citep{Heitmann_etal_2009,EuclidEmulator_etal_2018}. 


\section{Methodology}\label{sec:method}

In this Section, we describe the basics of Gaussian process emulation of cosmological statistics. We also detail our pipeline, which starts from the linear matter power spectrum of an arbitrary cosmology and maps this onto a point in the multi-dimensional parameter space over which the emulator performs the interpolation. This gives the emulator high versatility, in that it will be able to make predictions for a broad range of non-standard cosmologies absent from the training set.

	\subsection{Emulation strategy} \label{subsec:GPR}

\begin{figure}
\centering
\includegraphics[width=0.5\textwidth]{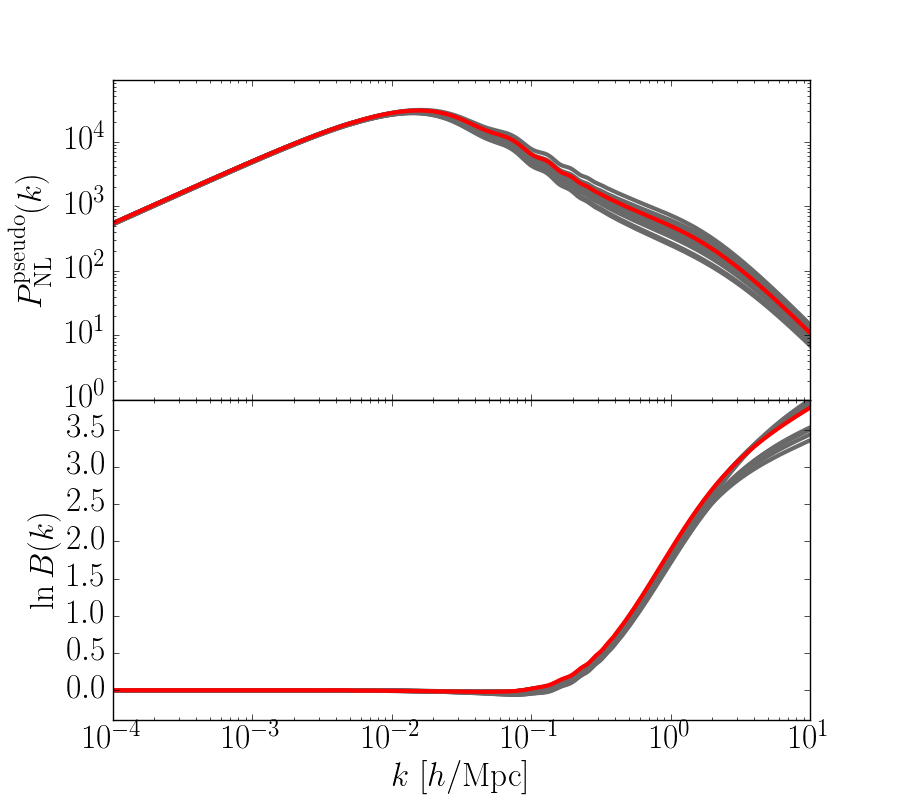} 
\caption{\textit{Upper panel:} in grey, the $z=0$ pseudo non-linear matter power spectra for the extensions to $\Lambda$CDM listed in Table \ref{tab:PhysMod_Table}, computed with {\sc{halofit}} as discussed in Section \ref{subsec:C19}. \textit{Lower panel:} the natural logarithm of the boost factor obtained by taking the ratio of pseudo non-linear and linear matter power spectra (see equation \ref{eqn:Boost}). For reference, the base $\Lambda$CDM cosmology (A), given in Table \ref{tab:PhysMod_Table}, is shown in red. The monotonicity and narrow dynamic range of the $\ln B$ relative to the $P^{\rm{pseudo}}_{\rm{NL}}$ help to improve the emulator's performance significantly.} \label{fig:Boostzdep}
\end{figure}

An in-depth discussion of our emulation strategy, employing Gaussian process (GP) regression, can be found in Appendix \ref{Appendix:GPR}. The accuracy of this method is sensitive to the particular observable being emulated. Our aim is to predict the \textit{pseudo} non-linear matter power spectrum discussed in Section \ref{subsec:C19}, a quantity which has a dynamic range of many orders of magnitude and features a change in gradient sign between small and large scales, as shown in the upper panel of Figure \ref{fig:Boostzdep} for the extensions to $\Lambda$CDM listed in Table \ref{tab:PhysMod_Table}. GP emulators on the other hand perform better at predicting smooth monotonic functions with a narrow dynamic range. Hence, following \cite{EuclidEmulator_etal_2018}, we consider the natural logarithm of the \textit{boost factor}, $B(k,z)$, as our observable of interest, where

\begin{equation}
B(k,z) \equiv \frac{P^{\rm{pseudo}}_{\rm{NL}}(k,z)}{P^{\rm{pseudo}}_{\rm{L}}(k,z)} . \label{eqn:Boost}
\end{equation} 

\noindent The lower panel of Figure \ref{fig:Boostzdep} shows that the natural logarithm of the boost factor indeed matches the desired properties of a GP observable. 

An essential step in the emulator design is data compression. For any particular cosmological model, the power spectrum is sampled both in spatial frequencies (wavenumbers) and time (redshifts), for a total of $\sim 10^4$ support points. Training the emulator on each of these individual measurements requires prohibitively large computational resources. To overcome this issue, we perform a principal component analysis (PCA) following previous work \citep[see, e.g,][]{Habib_etal_2007,Schneider_etal_2008,Heitmann_etal_2016}. For a given parameter vector, $\boldsymbol{\pi}$, and redshift, $z$, we apply the following linear decomposition

\begin{equation}
\ln B(k,z; \boldsymbol{\pi}) = \mu(k,z) + \sum_{j=1}^{n_{\psi}} \psi_j(k,z) w_j(\boldsymbol{\pi}) + \epsilon \,,
\label{eqn:Boost_PCA}
\end{equation}

\noindent where $\mu$ and $\{ \psi_1, \cdots,  \psi_{n_{\psi}} \}$ are, respectively, the mean and basis functions of the training set of boosts at $z$, and the weights $w_j$ are the projections of the mean-subtracted observable onto $\psi_i$. The error term $\epsilon$ has two contributions: one component, arising from the negligible noise on the {\sc{halofit}} training predictions (given by $\epsilon_{\rm n}(\boldsymbol{\pi})$ in equation \ref{eqn:yModel}); and a second component coming from missing relevant basis functions. We find that $n_\psi=12$ is sufficient to satisfactorily reconstruct the training set, and including more basis functions has only a marginal effect on the emulator accuracy. Ultimately, for a given $\boldsymbol{\pi}^\ast$ our emulator infers the boost factor by predicting the weights, $w_j(\boldsymbol{\pi}^\ast)$, conditioned on the training set $w_j(\boldsymbol{\pi})$.

	\subsection{Model-independent parameterisation of the pseudo cosmologies} \label{subsec:MIP}

Extensions to $\Lambda$CDM are described by an arbitrary number of additional physical parameters. Consider, for example, $w$CDM, $f(R)$ gravity and massive neutrino cosmologies: matter clustering in all these models depends on five standard cosmological parameters, e.g. $\boldsymbol{\pi}_\Lambda = \{ \Omega_{\rm{b}}h^2,\Omega_{\rm{m}}h^2,h,n_s,\ln(10^{10}A_s) \}$, along with one or more parameters describing model-specific features, $\boldsymbol{\pi}_{\rm X} = \{ f_{R0}, \sum m_\nu, \cdots \}$. To construct an emulator capable of predicting generic pseudo non-linear matter power spectra, we opt for a model-independent parameterisation mapping an arbitrary model to a certain coordinate, $\boldsymbol{\pi^*} = \{ \boldsymbol{\pi}_\Lambda, \Delta\boldsymbol{\alpha} \}$, in our emulation parameter space, where $\Delta\boldsymbol{\alpha}$ is a vector of effective parameters that quantifies smooth deviations from the linear matter power spectrum of the standard cosmology. Thus, for $\Delta\boldsymbol{\alpha} = 0$ we recover the late-time physics of the vanilla \LCDM{} model. For this, we model the `shape', namely the ratio of the \textit{linear} pseudo matter power spectrum to that of the \LCDM{} cosmology sharing the same standard parameters, as

\begin{multline}
\begin{split}
S(k,z;\boldsymbol{\pi}_{\Lambda},\Delta\boldsymbol{\alpha}) & \equiv \frac{P^{\rm{pseudo}}_{\rm{L}}(k,z;\boldsymbol{\pi}_{\Lambda},\boldsymbol{\pi}_{\rm X})}{P_{\rm L}^\Lambda(k,z;\boldsymbol{\pi}_{\Lambda})} \\
& \approx 1 + \sum_{i=1}^{n_{\Phi}} \Phi_i(k,z) \Delta\alpha_i \,,
\end{split}
\label{eqn:Shape}
\end{multline}

\noindent where $\{ \Phi_1, \cdots,  \Phi_{n_{\Phi}} \}$ are the principal components of a training set, $\mathbb{T}$, consisting of smooth curves capturing a range of `reasonable' deviations from $\Lambda$CDM~\footnote{Eq.~\eqref{eqn:Shape} can be understood in terms of a principal component reconstruction based on the training set $\mathbb{T}$, that is
\begin{equation}
S(k_m) \approx \nu(k_m) + \sum_{i=1}^{n_{\Phi}} \Phi_i(k_m)\alpha_i \,,
\label{eqn:Shape2}
\end{equation} 
where for simplicity we dropped the dependence on redshift and cosmology, and made explicit that we sample the shape $S$ at $k_m \in \{ k_1, \cdots,  k_{n_k} \}$. The function $\nu$ is the mean across $\mathbb{T}$, and in general we expect $\nu \neq 1$. Then, for any \LCDM{} cosmology we must have 
\begin{equation}
1 - \nu(k_m) \approx \sum_{i=1}^{n_{\Phi}} \Phi_i(k_m)\alpha_i^\Lambda \,,
\label{eqn:ShapeLCDM}
\end{equation}
with $\alpha_i^\Lambda$ being the weights that exactly compensate the departure of the mean from unity, and upon using the orthonormality of the 
basis functions they read
\begin{equation}
\alpha_i^\Lambda = \sum_{m=1}^{n_k} \Phi_i(k_m)[1 - \nu(k_m)]  \,.
\end{equation}
Finally, the combination of Eqs.~\eqref{eqn:Shape2} and~\eqref{eqn:ShapeLCDM} gives Eq.~\eqref{eqn:Shape}, with $\Delta\alpha_i \equiv \alpha_i - \alpha_i^\Lambda$. Note that for a given training set, $\mathbb{T}$, the \LCDM{} weights need to be computed only once, and can subsequently be hard-coded in the emulator.}. Use of PCA to parameterise deviations from vanilla $\Lambda$CDM cosmology has been previously implemented in a number of studies \citep[see, e.g.,][]{Zhao_etal_2009, Hojjati_etal_2012, Eifler_etal_2015}.

		\subsubsection{Generation of the basis set} \label{subsubsec:Random_Curves}
		
	We assemble a set of orthogonal basis functions to reconstruct a broad class of scale-dependent linear departures from the \LCDM{} power spectrum. For this purpose, we generate random curves in Fourier space over the range $10^{-4} < k \,{\rm Mpc}/h <10$, with the constraint that all curves must converge to unity on large enough scales. This last requirement is motivated by the physics of well-known \LCDM{} extensions, such as Generalised Brans-Dicke theories~\citep{De_Felice_Tsujikawa_2010,Park_et_al_2010,Hinterbichler_et_al_2011,Pogosian_Silvestri_2016,Quiros_et_al_2016,Joyce_et_al_2016} and massive neutrino cosmologies~\citep{Lesgourgues_Pastor_2006}. Both these models possess a characteristic scale, $\bar k$, such that on scales $k \ll \bar k$ the linear matter clustering matches that of the vanilla \LCDM{} cosmology. For $k \gg \bar k$ the growth of structure is either suppressed ($S < 1$) or enhanced ($S > 1$). Note that models with $\bar k \rightarrow 0 $ exhibit in practice a scale-independent linear growth~\citep[see, e.g.,][]{Pogosian_Silvestri_2016}, and a simple rescaling of the amplitude of the vanilla \LCDM{} power spectrum is sufficient to produce the corresponding pseudo power spectrum. Viable models in these two classes of theories above have $\bar k \gtrsim 10^{-3} \, h/\rm{Mpc}$~\citep{Brax_et_al_2012,Wang_et_al_2012,Joyce_et_al_2015,Planck_2018}. We therefore adopt the value $\tilde{k} =10^{-3} \, h/\rm{Mpc}$ for the scale at which the shapes converge to unity, and allow for generous positive smooth deviations in the range $k \in [\tilde{k}, 10] \, h/{\rm Mpc}$. Moreover, since the basis functions derived from our generated shapes are sufficiently general to describe the power spectrum ratios at various redshifts, we drop the $z$-dependence, $\Phi_i(k,z) \rightarrow \Phi_i(k)$, and use one basis set for all redshifts.

In order to reduce the sensitivity of the emulator to the particular method used to generate the set of random shapes, we employ two different and independent generating strategies. The first of these methods, which we refer to as {\sc{GPCurves}}, draws shapes from a 1-dimensional Gaussian process with a squared exponential kernel of the form given in equation \ref{eqn:kernel} (where $d=1$), conditioned on a `training set' of densely sampled points at $k<10^{-3} \, h/\rm{Mpc}$, all with values equal to unity, helping to fix the random shapes to this value on large scales. Here, for the kernel amplitude, we use $A=5$, which corresponds to the maximum value that the random shapes can have on scales $k>10^{-3} \, h/\rm{Mpc}$. We tune the only length-scale parameter, $p$, so that the artificial shapes exhibit at most one stationary point for $k>10^{-3} \, h/\rm{Mpc}$, thus resembling features of the physical shapes in Figure \ref{fig:PhysicalShapes}. There is no intrinsic restriction in the GP preventing generated shapes from taking unphysical negative values. We therefore generate an initial set of curves from which we sample 1000 viable positive shapes.

As a second method we use {\sc{Smurves}} \citep{Moews_etal_2019}, a novel random curve generator. The functionality of the generator is illustrated by considering a particle travelling in Fourier space from low-to-high wavenumbers. Before $k=10^{-3} \, h/\rm{Mpc}$, the particle experiences no forces and hence travels undisturbed at a vertical coordinate of unity. After this point, forces randomly generated from a uniform distribution perturb the particle in the vertical direction, altering its trajectory. Also randomly generated are the locations of gradient sign changes. In this manner, we produce an additional 1000 smooth curves with a maximum of one stationary point and vertical intervals in the range [0, 8]. In order to facilitate extra flexibility in the deviations from $\Lambda$CDM, the parameters specifying the curve generation in {\sc{Smurves}} are set purposefully to create a sample of curves which are slightly broader than {\sc{GPCurves}}.

\begin{figure}
\begin{center}
\includegraphics[width=\columnwidth]{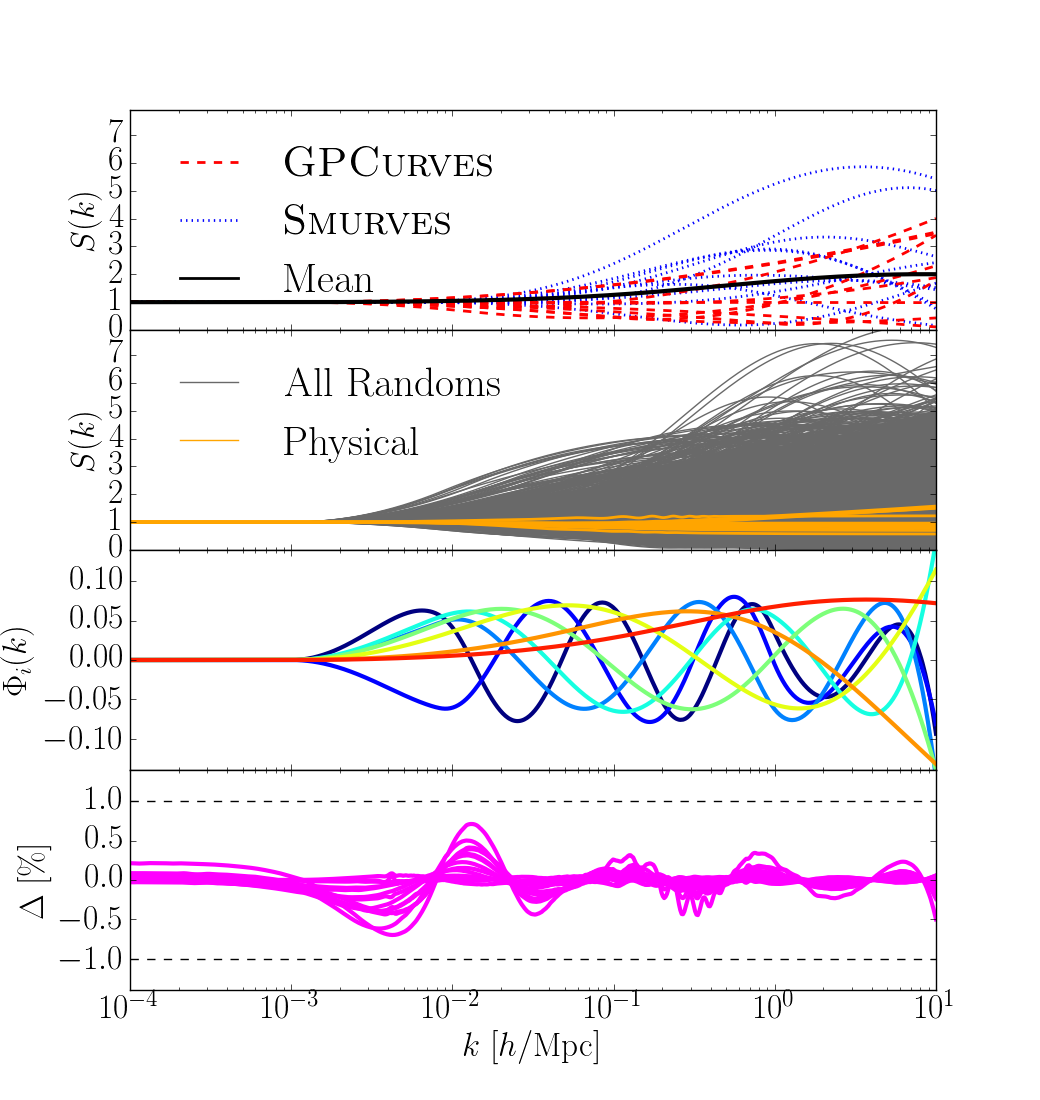}
\caption{\textit{Upper panel:} A sub-sample of random shapes generated with the {\sc{GPCurves}} (dashed red) and with the {\sc{Smurves}} (dotted blue) methods. The mean across all 2000 random shapes is shown in solid black. \textit{Upper-middle panel:} All 2000 random shapes from {\sc{GPCurves}} and {\sc{Smurves}} (grey) with the 14 shapes for the physical models from Figure \ref{fig:PhysicalShapes} shown for reference (orange). \textit{Lower-middle panel:} the 8 orthogonal basis functions, $\Phi_i$, obtained via a PCA of the 2000 random shapes. The colour indicates the ranking of the basis function in the hierarchy of explaining the variance in the shapes, with red indicating the most important, and dark blue indicating the least. \textit{Lower panel:} the accuracy, $\Delta$, in reconstructing the physical model shapes, from Figure \ref{fig:PhysicalShapes}, using the basis functions, $\Phi_i$. } \label{fig:Rand_Curves}
\end{center}
\end{figure}

A comparison of 10 curves randomly selected from either samples are shown in different colours in the upper panel of Figure \ref{fig:Rand_Curves}. The upper-middle panel shows the shapes corresponding to the physical models in Table \ref{tab:PhysMod_Table} (orange) relative to the collective sample of 2000 random curves from {\sc{GPCurves}} and {\sc{Smurves}} (grey). The lower-middle panel shows the orthogonal basis functions, $\Phi_i$, obtained via a PCA of the 2000 curves. Note that the resulting basis set allows for a broader range of departures from $\Lambda$CDM than our test physical models.

Each basis function, $\Phi_i$, appearing in the shape reconstruction (equation \ref{eqn:Shape}) contributes one extra dimension to the parameter space of the emulator, with the additional parameters corresponding to the PCA weights, $\Delta\alpha_i$. Therefore, in order for our emulator to be computationally efficient whilst still achieving accuracies of $\lesssim$1\%, we keep the smallest number of basis functions that guarantees sub-per-cent reconstruction errors on all physical test shapes (see Figure \ref{fig:PhysicalShapes}). We find that $n_\Phi=8$ basis functions are sufficient to reconstruct 
$f(R)$ gravity shapes with negligible errors. For cosmologies with massive neutrinos, however, changes to the BAO dynamics at early times produce rapid oscillations in their late-time power spectrum shapes. These features require one additional step in the reconstruction, but no extra parameters need be included in the emulation. Further details on this can be found in Appendix \ref{Appendix:BAO_Modelling}. The resultant shape reconstruction accuracy for all physical models is shown in the lower panel of Figure \ref{fig:Rand_Curves}. Overall our emulation volume is therefore carved out of a 13-dimensional space: 5 \LCDM{} parameters, $\boldsymbol{\pi}_\Lambda$, and 8 shape parameters, $\Delta\boldsymbol{\alpha}$.

	\subsection{Building the training and trial sets} \label{subsec:BuildingTrainTrial}

Having established a model-independent description for arbitrary cosmological models, we now discuss how we generate training and trial sets for our pseudo matter power spectrum emulator. In Section \ref{subsubsec:LHC} we give details on the Latin hypercube (LH) space-filling strategies we adopt to efficiently sample our 13-dimensional space. In Section \ref{subsubsec:Param_Range}, we then define suitable ranges for the cosmological and shape parameters, $\{ \boldsymbol{\pi}_{\Lambda}, \Delta\boldsymbol{\alpha} \}$, such that the emulated boost factors are representative of viable extensions to $\Lambda$CDM. The nodes of the LHs take values in the range $[0,1]$, which we then properly rescale to obtain the ``physical'' coordinates $\boldsymbol{\pi}$ used to generate the training and trial sets.

	\subsubsection{Latin hypercube sampling} \label{subsubsec:LHC}

	The use of LH-based distributions of points to evenly sample a parameter space was first introduced by \citet{McKay_etal_1979}. The LH is a generalisation of the concept of a Latin square in which each row and column of a two-dimensional grid features exactly one sample, taking the form of a chess board with a number of rooks that do not threaten each other. Owing to the desirable space-filling properties, LH sampling of the input parameters for simulations has become a standard practice \citep{Sacks_etal_1989,Currin_etal_1991,Heitmann_etal_2006, Schneider_etal_2008, Agarwal_etal_2012, Liu_etal_2015, Garrison_etal_2018, Harnois-Deraps_etal_2019}.

Assembling 13-dimensional LH designs therefore presents an ideal way to construct predictions to both train the emulator and test its accuracy for different regions of the emulation space. The spatial distribution of the trial cosmologies does not require optimisation. Therefore we sample 300 trial coordinates from a 13-dimensional LH generated with the {\sc{pyDOE}} Python package\footnote{\url{https://pythonhosted.org/pyDOE/}}. The training nodes, however, impact significantly on the emulator accuracy, and sampling methods able to cover the high-dimensional volume as uniformly as possible are thus preferable. For this purpose, space-filling criteria based on the distance between samples are commonly used. One of the most prominent approaches is the Maximin LH design, which maximises the minimum distance between samples \citep{Morris_Mitchell_1995}. While this approach leads to the sought-after space-filling properties in the bulk of the parameter space, most points cluster in the corners or at the boundaries of the lower-dimensional projection spaces. Since the matter power spectrum is most sensitive to subsets of cosmological and shape parameters, space-filling strategies also optimising projection space sampling would give an obvious advantage.

To this end, \citet{Joseph_etal_2015} introduced the maximum projection design ({\sc{MaxPro}}), in which a weighted Euclidean distance, $E$, is minimised, resulting in the new criterion

    \begin{equation}
    \min_{\boldsymbol{D}} E (\boldsymbol{D}) = \left( \frac{1}{{N \choose 2}} \sum_{i = 1}^{N - 1} \sum_{j = i + 1}^N \frac{1}{\prod_{l = 1}^d (x_{il} - x_{jl})^2}\right)^{\frac{1}{d}},
    \end{equation}
 
   \noindent where $\boldsymbol{D} = \{ \boldsymbol{x}_1, \dots \boldsymbol{x}_N \}$ is an $N \times d$ design  matrix containing $N$ samples occupying the $d$-dimensional parameter space. $x_{il}$ denotes the value of the $i$-th sample's coordinate in the $l$-th dimension. The {\sc{MaxPro}} criterion maximises the bulk and projection space-filling properties, since for any dimension $l$, if $x_{il} = x_{jl}$ and $i \neq j$, $E(\boldsymbol{D}) = \infty$. This ensures that in the design minimising $E (\boldsymbol{D})$ no two points can be close to each other in any of the projections.

  The MaxPro design optimisation implemented in this work is based on the simulated annealing algorithm described in \citet{Morris_Mitchell_1995}\footnote{\url{https://CRAN.R-project.org/package=MaxPro}}. This consists of selecting an initial configuration before searching for progressively better design choices by randomly perturbing the current design, keeping the new matrix should it reduce the cost function, $E(\boldsymbol{D})$, or else do so with a given probability. Finally this implementation arrives at a locally optimal {\sc{MaxPro}} design by ascending the gradient given by

    \begin{equation}
    \frac{\partial E^d (\boldsymbol{D})}{\partial x_{rs}} = \frac{2}{{N \choose 2}} \sum_{i \neq r} \frac{1}{(x_{is} - x_{rs})} \frac{1}{\prod_{l = 1}^d (x_{il} - x_{rl})^2}.
    \end{equation}

\noindent To investigate the performance of our emulator we generate sets of training nodes with sizes $N=100,300,500$ and 600, with the additional property of having an optimal space-filling pattern in each projection. Furthermore, to investigate the sensitivity of our results to the specific configuration of nodes in each training set, we produce 10 optimised designs of each size and measure the range of achieved accuracies across them.

		\subsubsection{Setting parameter ranges} \label{subsubsec:Param_Range}
		
We shall now map our dimensionless training and trial nodes belonging to the unit LHs onto our parameters, $\boldsymbol{\pi} \in \{ \boldsymbol{\pi}_{\Lambda}, \Delta\boldsymbol{\alpha} \}$. To do this, we must identify appropriate ranges for these parameters. We begin with the $\Lambda$CDM parameters, and note that our emulator is intended for cosmological analyses employing high-quality data from the next-generation galaxy surveys. Hence, we set the boundaries on $\boldsymbol{\pi}_\Lambda$ to be roughly consistent with the 95\% marginalised constraints from the combined TT, TE, EE $+$ lowE analysis in \citet{Planck_2018}, namely,
\begin{equation}
  \begin{gathered}
 0.0215 < \Omega_{\rm{b}}h^2 < 0.0235, \\
 0.12 < \Omega_{\rm{m}}h^2 < 0.155, \\
 0.6 < h < 0.8, \\
 0.9 < n_s < 1.05, \\
 2.92< \log(10^{10}A_s) < 3.16. \\
  \end{gathered}
\end{equation}

\begin{figure}
\begin{center}
\includegraphics[width=0.5\textwidth]{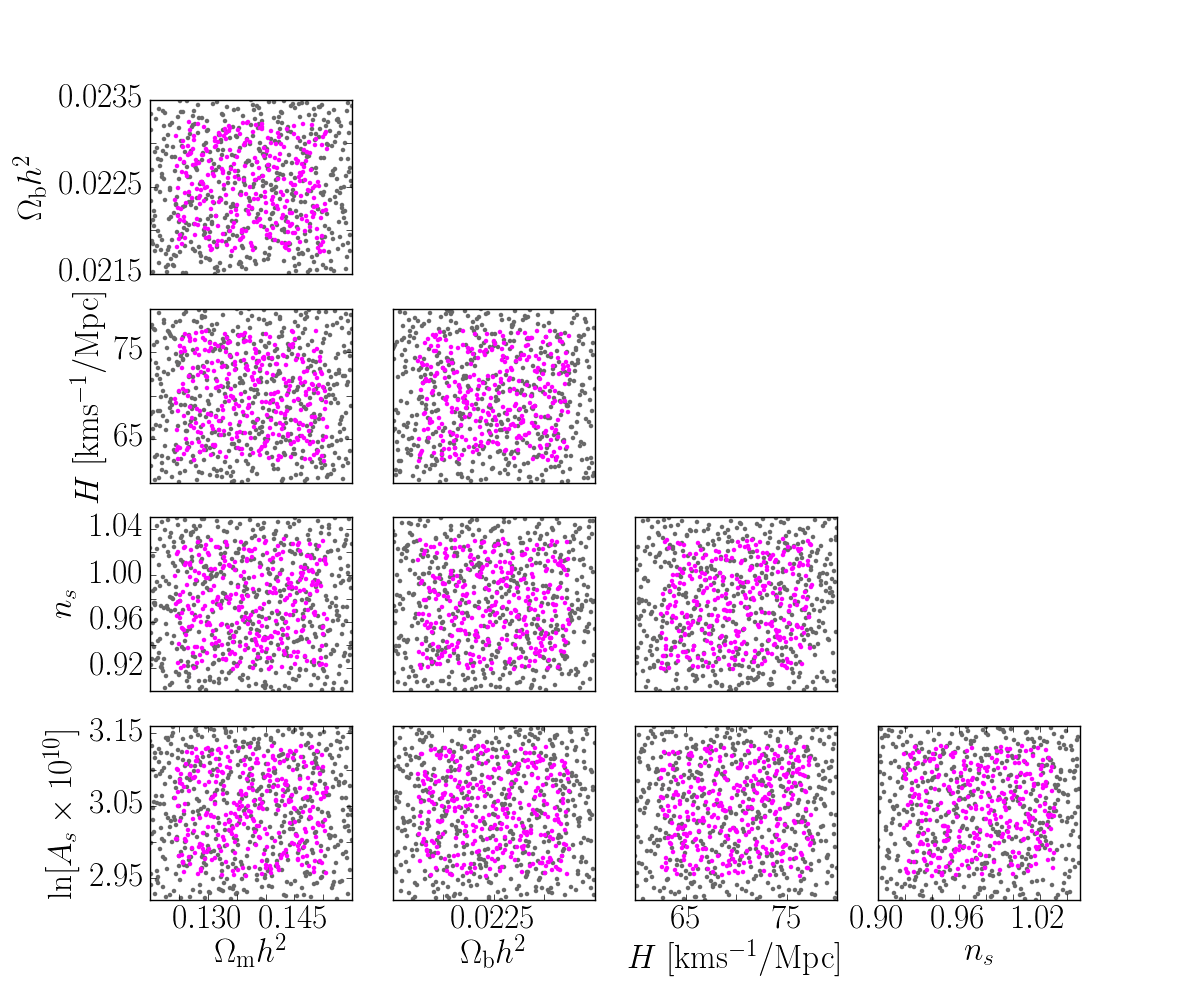} \\
\caption{The 2D projections of the 13-dimensional emulation space for the five $\Lambda$CDM parameters only. The grey points correspond to a 500-node training set obtained by maximising the distance between nodes in each projection, whereas the magenta points correspond to the trial coordinates, which are simply sampled from a standard Latin hypercube (see Section \ref{subsubsec:LHC} for more details). The axes show the full allowed range of the training points. The trial coordinates are confined to a smaller hypercube with sides measuring 75\% of the training range, which reduces the impact of boundary effects on the emulator accuracy. } \label{fig:TrainingSetCosmol}
\end{center}
\end{figure}

\noindent These ranges are illustrated in Figure \ref{fig:TrainingSetCosmol}, which shows the 2D projections of the $\Lambda$CDM sub-space, with the grey points corresponding to one of the $N=500$ {\sc{MaxPro}} distributed training designs.

\begin{figure}
\begin{center}
\includegraphics[width=0.5\textwidth]{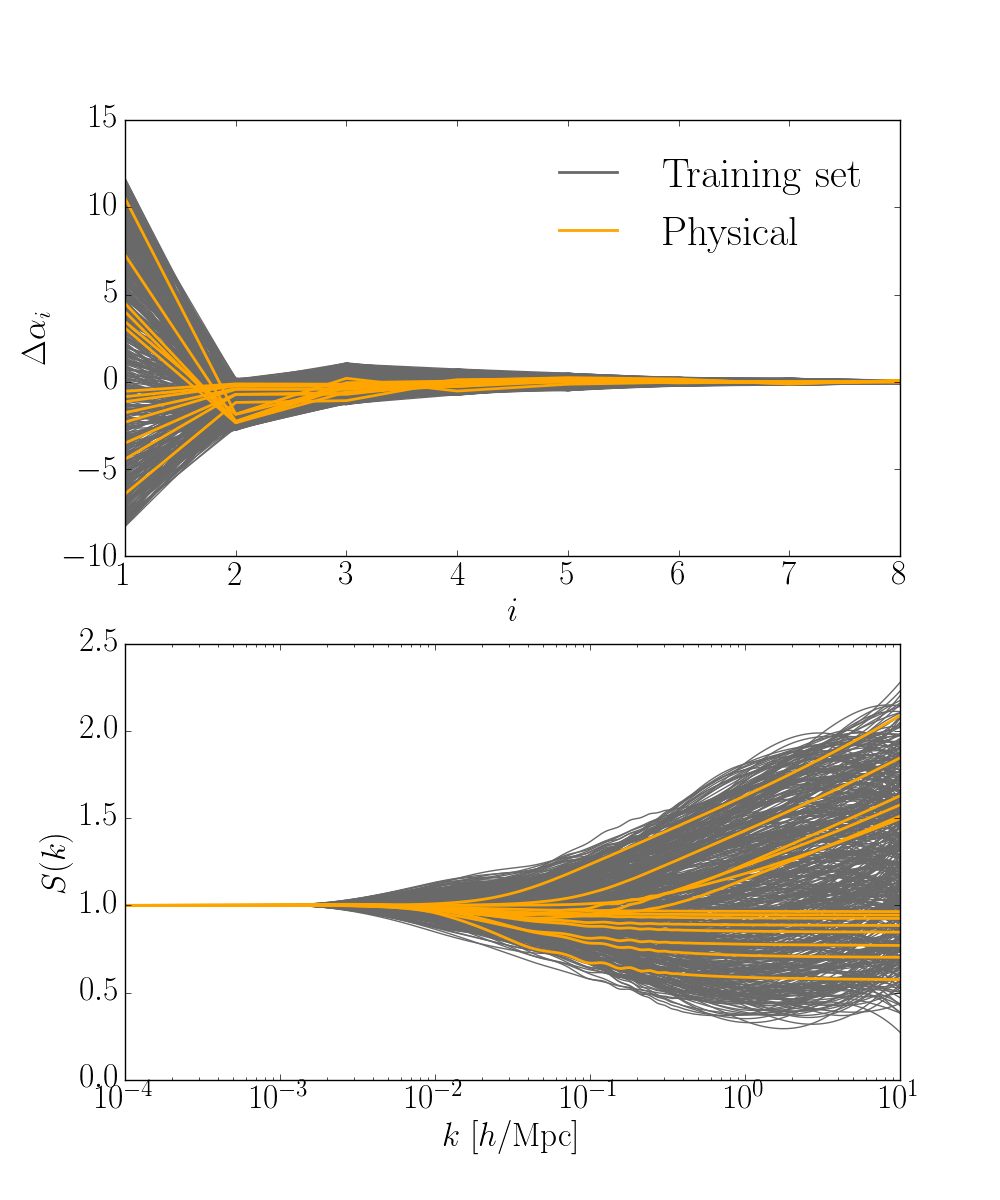} \\
\caption{\textit{Upper panel}: the $\Delta\alpha_i$ values, as a function of PCA rank, $i$, which parameterise deviations from \LCDM{} in the $z=0$ pseudo linear matter power spectra, for a 500 node training set (grey) and physical models (orange). \textit{Lower panel}: synthetic and physical shapes corresponding to the $\Delta\alpha_i$ weights above.} \label{fig:TrainingSetShapes}
\end{center}
\end{figure}

Turning to the shape parameters, $\Delta\boldsymbol{\alpha}$, we take inspiration from the physical models listed in Table \ref{tab:PhysMod_Table}, and simply set the range of values in our training set to be slightly broader than those defined by the $\Delta\boldsymbol{\alpha}$'s associated with $f(R)$ gravity, massive neutrino and combined cosmologies. The upper panel of Figure \ref{fig:TrainingSetShapes} shows the $\Delta\alpha_i$ for the physical models at $z=0$ in orange, and those related to one of the designs with 500 optimally spaced training nodes (see Section \ref{subsubsec:LHC}) in grey. The convergence of the $\Delta\alpha_i$ values towards zero as the PCA rank, $i$, increases, is an indication that we are using a  sufficient number of basis functions to reconstruct our test models.

The lower panel of Figure \ref{fig:TrainingSetShapes} shows the shapes corresponding to the $\Delta\boldsymbol{\alpha}$ vectors shown in the upper panel, both for the 500-node training set (in grey) and for the physical models (in orange), reproduced from Figure \ref{fig:PhysicalShapes}. We see that the training shapes cover a complex range of behaviours, whilst still encompassing the shapes of the physical models.

For the trial set, we opt for narrower parameter ranges, such that in each dimension the trial coordinates have values in the central 75\% of the training set intervals. In doing this we test the emulator accuracy only in the bulk of the parameter space, so that our conclusions are less influenced by edge effects, where emulator accuracy degrades due to a shortage of nodes near the boundaries. This is visualised in Figure \ref{fig:TrainingSetCosmol} by the distribution of trial nodes (in magenta). We also test the performance of our emulator for strictly \LCDM{} cosmologies, where all shape parameters are set to zero. This $\Lambda$CDM trial ensemble is used to test whether the emulator is able to make accurate predictions for the standard model in the non-linear regime, even when $\Lambda$CDM cosmologies are completely absent from the training set.

\subsubsection{Generating boost factors}

\begin{figure}
\begin{center}
\includegraphics[width=0.5\textwidth]{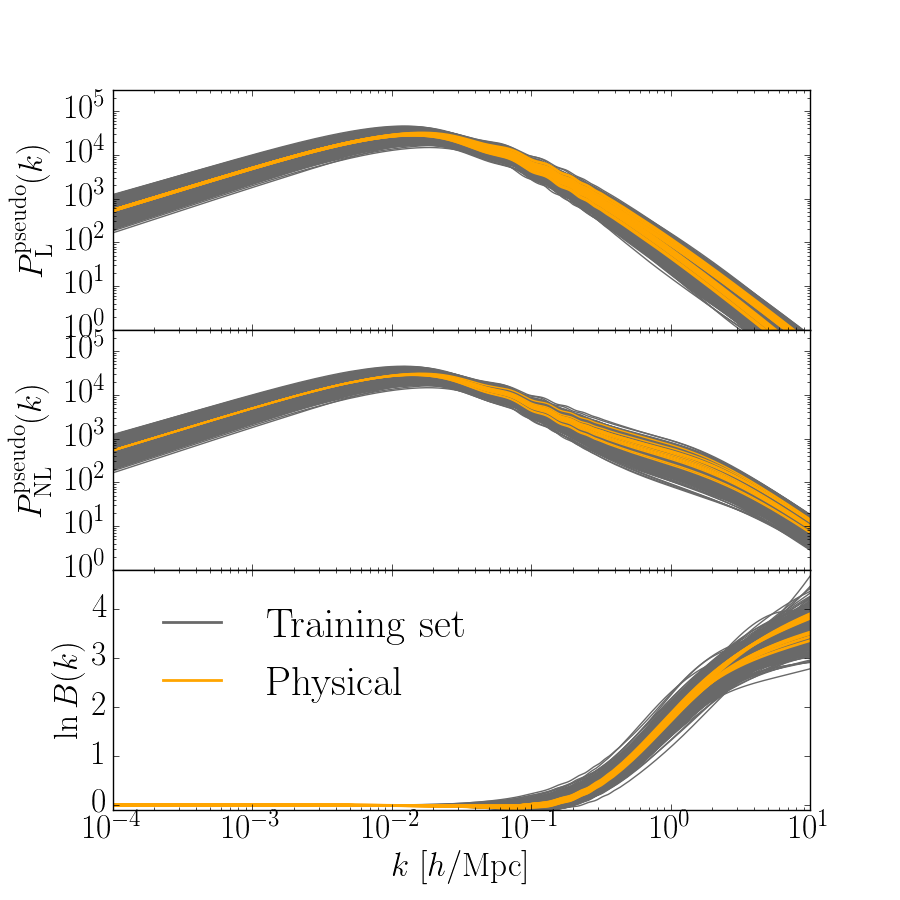} \\
\caption{The $z=0$ linear (upper) and non-linear (middle) matter power spectra for the 500 pseudo cosmologies of a training set realisation (grey) and for the physical models (orange). The lower panel shows the logarithm of the boost factor, defined as the natural logarithm of the ratio of the middle to the upper panels (see equation \ref{eqn:Boost}). By design our training set comfortably encompasses the physical model predictions.} \label{fig:TrainingSetPk}
\end{center}
\end{figure}

With the training and trial nodes now assembled, the final task is to produce the corresponding linear as well as non-linear matter power spectra, and hence the logarithm of the boost factor, $B(k,z)$ (see equation \ref{eqn:Boost}), which the emulator is trained on and predicts. Conventionally, the non-linear statistics are measured from cosmological simulations, but for the purposes of this proof-of-concept study {\sc{halofit}} can serve as a proxy for a suite of $N$-body simulations. Note that this approach is self-consistent as long as the growth of structure and background evolution of the `simulated' cosmologies matches those of the \LCDM{} model, which is indeed the case for the pseudo cosmologies. To estimate the surrogate $P^{\rm{pseudo}}_{\rm{NL}}$ we repeat the process described in Section \ref{subsec:C19} for the physical models: for a given pseudo cosmology, $\boldsymbol{\pi} = \{ \boldsymbol{\pi}_\Lambda, \Delta \boldsymbol{\alpha} \}$, our modified version of {\sc{camb}} accepts as input the random shape described by $\Delta \boldsymbol{\alpha}$, and internally multiplies this by the \LCDM{} linear power spectrum associated with $\boldsymbol{\pi}_\Lambda$. This generates the pseudo linear power spectrum which together with $\boldsymbol{\pi}_\Lambda$ serves as input for {\sc{halofit}} non-linear predictions. The boost factor is then produced by simply taking the ratio of the non-linear to the linear power spectrum. We generate boost factors for the training and trial sets of our emulator at redshifts 0 and 1. The $z=0$ pseudo linear and non-linear matter power spectra, along with the corresponding boost factors, are shown in Figure \ref{fig:TrainingSetPk} for a 500-node training set (in grey), as well as for the physical models (in orange) also illustrated in Figure \ref{fig:Boostzdep}.

\begin{figure*}
\begin{center}
\includegraphics[width=0.49\textwidth]{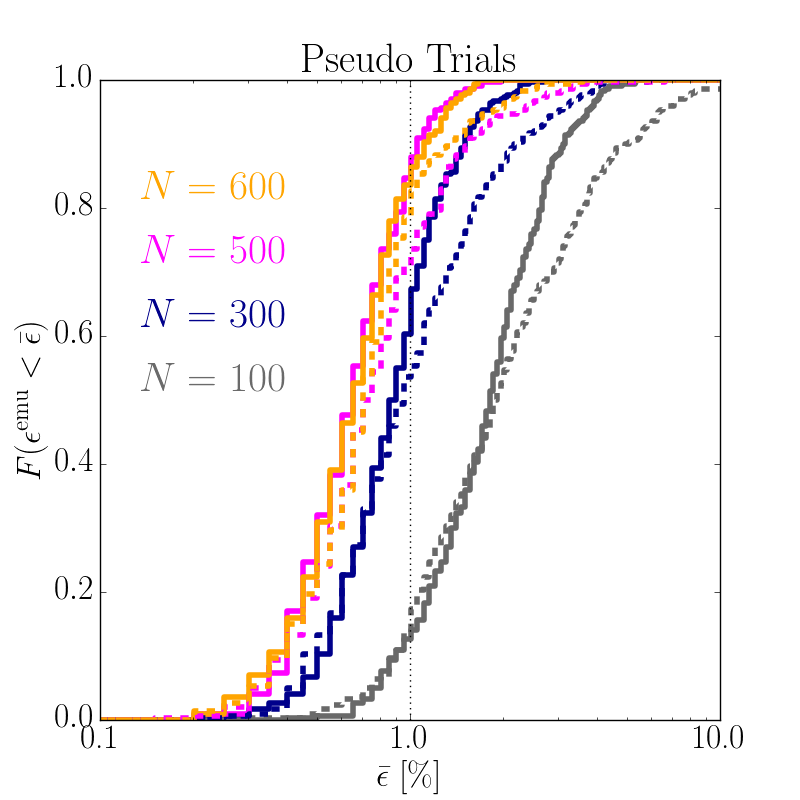} 
\includegraphics[width=0.49\textwidth]{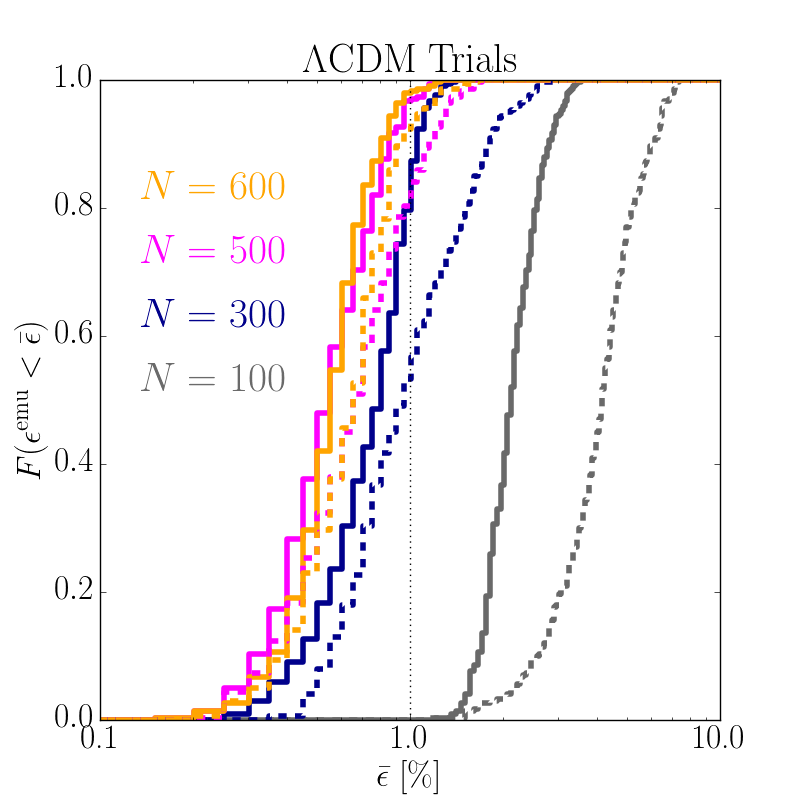} \\
\caption{The cumulative distribution, $F$, of trial predictions for which $\epsilon^{\rm emu}$, the maximum absolute fractional deviation between the emulated and {\sc{halofit}} pseudo non-linear matter power spectra over the range $k \in [0.01,10] \, h/\rm{Mpc}$ (equation \ref{eqn:epsilon_emu_max}), is within a threshold, $\bar \epsilon$, for $z=0$ (solid lines) and $z=1$ (dashed lines). $F$ is averaged across the 10 realisations per training set size, $N$, increasing from 100 nodes (grey) to 600 (orange). The left panel corresponds to the 300 pseudo cosmology trials. The right panel illustrates the performance for the 300 pure $\Lambda$CDM trials (see Section \ref{subsubsec:Param_Range}).} \label{fig:Bulk_Accuracy}
\end{center}
\end{figure*}


\section{Results} \label{sec:Results}

The non-linear matter power spectrum is key to the derivation of a number of cosmological observables, therefore we use this rather than the emulated statistic, $\ln B(k,z)$, to evaluate the accuracy of our emulator. We consider wavenumbers in the range $k \in [0.01, 10] \, h/\rm{Mpc}$, excluding larger scales where the boost factor goes to unity in all models, presenting no challenge to the emulator. For each trial cosmology and redshift separately, we quantify the emulator accuracy as

\begin{equation}
\epsilon^{\rm emu} \equiv \max_{0.01< k < 10} \left| \frac{ P^{\rm emu}_{\rm NL}(k) - P^{\rm True }_{\rm NL}(k) }{P^{\rm True }_{\rm NL}(k)} \right|
\label{eqn:epsilon_emu_max}
\end{equation}

\noindent where $P^{\rm emu}_{\rm NL}$ and $P^{\rm True}_{\rm NL}$ refer to the emulated and true pseudo non-linear power spectra respectively, and for the purposes of this study we take {\sc{halofit}} non-linear predictions as the truth. Figure \ref{fig:Bulk_Accuracy} presents the fraction of the trial cosmologies with emulation errors smaller than a threshold value, $\bar \epsilon$, for various sizes of the training sets at both $z=0$ (solid lines) and $z=1$ (dashed). The fraction shown is the average across the 10 realisations per training set size. The left panel shows the results for the pseudo cosmology trials, whilst the right panel corresponds to the pure $\Lambda$CDM trial ensemble.

At $z=0$, for the pseudo cosmology trials and training sets of size $N=500$ or 600, the emulation is accurate to better than 1\% for about $85\%$ of the trials, and the full ensemble is recovered to better than 2\% accuracy. As expected, the emulator performs noticeably worse when trained on a smaller number of nodes, with only 13\% (60\%) of the pseudo non-linear spectra emulated to better than 1\% when $N=100$ (300). This is because the emulator has less information on the complex relationship between input parameters, $\boldsymbol{\pi}$, and the boost factors. However, we find that the largest inaccuracy across all trial models is still $<6\%$ ($<3\%$) for $N=100$ (300).

At $z=1$, the fraction of pseudo trials with $\epsilon^{\rm emu}<1\%$ are similar to those at $z=0$. However, the reduced steepness of the cumulative distributions at $z=1$, compared to those at $z=0$, suggests an increasing fraction of outliers with redshift. This is caused by the training and trial boost factors displaying a broader dynamic range at the higher redshift, which poses a greater challenge to emulation. Such broadening can be understood in terms of differences in the boost factor evolution induced by the different growth rates and non-linear mode couplings for the various base \LCDM{} parameters and shapes.

The right panel of Figure \ref{fig:Bulk_Accuracy} shows similar trends for the \LCDM{} trial ensemble. For the two largest training sets we find that over 90\% of the trial cosmologies are reproduced with better than percent accuracy at both redshifts analysed. The larger emulation space dimensionality compared to similar emulator schemes designed for $w$CDM cosmologies only \citep[see, e.g.][]{Heitmann_etal_2014} demands a three-fold increase in the number of nodes to achieve equivalent accuracies for \LCDM{} cosmologies. In fact, for our smallest training set ($N=100$) the emulator predictions can have errors as large as 4\% (7\%) at $z=0$ ($z=1$), comparable to those obtained by \cite{Heitmann_etal_2014}. Although our emulation strategy requires a few hundred training nodes to reach percent accuracy, it is important to realise that it provides non-linear power spectrum predictions for a much broader class of cosmologies, including the \LCDM{} model. Moreover, with the aid of {\it pairing and fixing} techniques \citep{Angulo_Pontzen_2016,EuclidEmulator_etal_2018} the total number of training simulations required is analogous to that of other GP emulators \citep{Lawrence_etal_2010,Lawrence_etal_2017}.

\begin{figure}
\begin{center}
\includegraphics[width=0.5\textwidth]{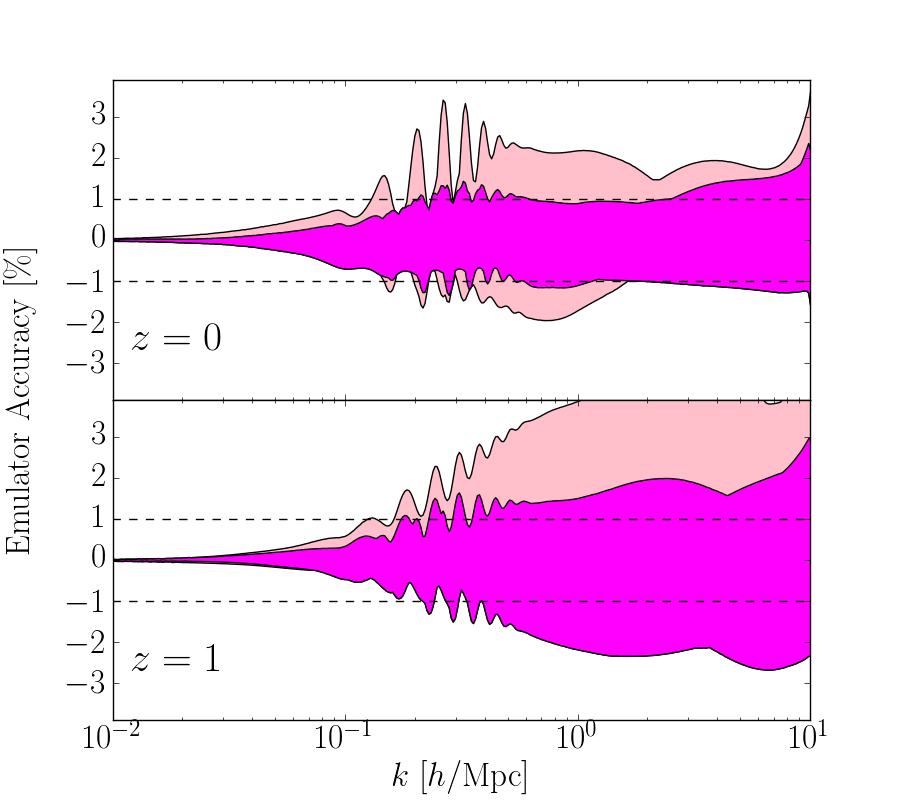} \\
\caption{The full range of accuracies of the emulated pseudo matter power spectra for the physical models at $z=0$ (upper panel) and $z=1$ (lower panel), obtained with our 10 training sets of 500 nodes. The difference in the outer (pale) and inner (dark) band is whether the xMNU{\_}0.4 cosmology is included in calculating the range of accuracies. Hence, much of the observed error bar can be attributed to this extreme model, which lies close to the edge of our training sets.} \label{fig:Model_Accuracy}
\end{center}
\end{figure}

Finally we assess the accuracy of the emulator for the physical models in Table \ref{tab:PhysMod_Table}. Motivated by our results for the synthetic trial cosmologies, for this second test we use $N=500$ training nodes, since a larger training set yields no significant improvement in the emulator performance.

The full range of the fractional errors of the emulated $P^{\rm{pseudo}}_{\rm{NL}}$ relative to the {\sc{halofit}} predictions, across the 10 training set realisations and all physical models, are shown in Figure \ref{fig:Model_Accuracy} for $z=0$ (upper panel) and $z=1$ (lower panel). The outer (pale) and inner (dark) bands show the ranges including or excluding the xMNU{\_}0.4 cosmology, respectively. Comparison of these reveals that much of the observed error is associated with this extreme model, which resides outside of the central 75\%-per-side hypervolume used in evaluating the accuracies with the synthetic pseudo cosmologies. We find that the remaining models are recovered to better than the target percent accuracy for $k<10 \, h/\rm{Mpc}$ at $z=0$, with the majority of the training set realisations. We note that the range plotted, showing the total variation obtained across the training set realisations, presents a conservative estimate of the emulator performance; the $2\sigma$ range of \textit{absolute} emulation accuracies, for example, is considerably tighter. We observe a slightly higher outlier rate at $z=1$, consistent with our observations with the synthetic pseudo cosmologies. Of the models included in the dark band, we find, once again, that the largest inaccuracies occur for cosmologies near the boundary of the sub-volume used to test the synthetic shapes, MNU{\_}0.4 and xMNU{\_}0.2, both lying many standard deviations away from the {\it Planck} best fit cosmology \citep{Planck_2018}.


\section{Discussion and Conclusions} \label{sec:Conclusions}

In this paper, we have designed and tested a methodology essential for predicting accurate non-linear matter power spectra in a broad range of extensions to the \LCDM{} cosmology in the context of the \textit{halo model reaction} framework \citep[`C19']{Cataneo_etal_2019}. The effectiveness of the C19 strategy rests on the availability of accurate baseline \LCDM{} non-linear power spectra evolved from non-standard initial conditions, the \textit{pseudo} power spectra (see Section \ref{subsec:C19}). We showed that such quantities can be readily predicted with a Gaussian process emulator whilst simultaneously satisfying the stringent accuracy requirements. We demonstrated the power of this technique for theoretically motivated models, such as $f(R)$ gravity and massive neutrino cosmologies, as well as for phenomenological models, where in both cases structures on different scales grow at different rates. 

Based on the results of this work, we advocate the following implementation of our methodology in the future:

\begin{enumerate}

\item A 13-dimensional optimised Latin hypercube is defined in terms of: $\boldsymbol{\pi}_\Lambda \, (\rm{dim}=5)$, the cosmological parameters that define the linear \LCDM{} power spectrum, and $\Delta \boldsymbol{\alpha} \, (\rm{dim}=8)$, the shape parameters which modify the linear \LCDM{} power spectrum to create the linear power spectrum for a given beyond-\LCDM{} model.

\item Two standard \LCDM{} $N$-body simulations,  with  initial conditions \textit{paired} and \textit{fixed} following \cite{Angulo_Pontzen_2016}, are run for each node in the LH, i.e. with the input cosmology $\boldsymbol{\pi}_\Lambda$, and rescaling of the primordial power spectrum with the shape defined by $\Delta \boldsymbol{\alpha}$.

\item From the output of each $N$-body simulation, the pseudo non-linear matter power spectrum, $P^{\rm pseudo}_{\rm NL}(k,z)$, and hence the natural logarithm of the boost factor, $\ln B(k,z)$ (see equation \ref{eqn:Boost}), are measured. An emulator is subsequently trained to predict this quantity for any $\{ \boldsymbol{\pi}_\Lambda, \Delta \boldsymbol{\alpha} \}$ combination.

\item A user will provide the analytically computed linear power spectrum in their chosen beyond-\LCDM{} model, $P_{\rm L}^{\rm real}(k,z; \boldsymbol{\pi}_\Lambda, \boldsymbol{\pi}_X)$, where $\boldsymbol{\pi}_X$ is a vector of parameters describing the particular extension to the standard cosmology, and the corresponding vector of base cosmological parameters, $\boldsymbol{\pi}_\Lambda$. We then perform a PCA decomposition of the ratio, $P_{\rm L}^{\rm real}(k,z; \boldsymbol{\pi}_\Lambda, \boldsymbol{\pi}_X)/P_{\rm L}^\Lambda(k,z; \boldsymbol{\pi}_\Lambda)$, where the denominator is the standard \LCDM{} linear spectrum, to map $\boldsymbol{\pi}_X$ to the eight PCA weights (see equation \ref{eqn:Shape}).

\item Given the set of values of the base cosmological parameters and PCA weights for the queried model, the emulator predicts $\ln B(k,z)$.

\item The pseudo non-linear matter power spectrum is then simply obtained as $P_{\rm NL}^{\rm pseudo}(k,z) = B(k,z) \times P_{\rm L}^{\rm real}(k,z)$.

\item Finally, a rescaling of the pseudo non-linear power spectrum by the C19 halo model reaction, in equation \ref{eq:HMR_frame}, produces the \textit{real} non-linear power spectrum including the effects of physics for the chosen beyond-\LCDM{} model.

\end{enumerate}

\medskip

The advantage of our emulator scheme over existing approaches \citep{Heitmann_etal_2014,Lawrence_etal_2017, EuclidEmulator_etal_2018,Winther_etal_2019} is twofold: the model-independent parameterisation enables predictions for beyond-one-parameter extensions to \LCDM{}, effectively expanding the domain of applicability to a much broader class of cosmologically interesting models; and the simulations required for the training phase are all based on the standard \LCDM{} cosmology, which makes them easier and faster to run. 

This work is a first step towards a proper and accurate non-linear matter power spectrum emulator, one with a training set built from the output of $N$-body simulations. Here, instead, our goal was to perform a feasibility study, in which the approximate semi-analytical predictions of {\sc{halofit}} act as surrogate simulations (replacing step (ii) above). Taking inspiration from the linear power spectrum shape of well-studied extensions to the standard cosmology, we computed the training pseudo non-linear power spectra by feeding smooth, random modifications of the \LCDM{} linear power spectrum to {\sc halofit}. Despite its intrinsic 5-10\%  inaccuracies, previous studies showed that {\sc{halofit}} provides a quick and robust way to design, construct and test the emulator \citep{Heitmann_etal_2009,EuclidEmulator_etal_2018}.

With the aid of \textit{pairing and fixing} techniques for the initial conditions \citep{Angulo_Pontzen_2016}, we estimate that a total of 1000 \LCDM{}-\textit{like} simulations (two realisations for each of the 500 training nodes) is enough to build a per-cent-level accurate pseudo non-linear matter power spectrum emulator. This setup also allows for predictions of the $\Lambda$CDM non-linear matter power spectrum with $\lesssim 1\%$ accuracy, even with the standard cosmology being completely absent from the training set. Despite the remarkable amount of computational resources required, the number of simulations is not too dissimilar from that used for the training of previous, more limited emulators \citep{Lawrence_etal_2010,Lawrence_etal_2017}.

Although not shown here, our methodology can also potentially be extended to include the effects on baryonic physics on non-linear scales \citep[see, e.g.,][]{Schneider_etal_2019, Debackere_etal_2019}. This correction, which could be incorporated into the halo model reaction, is left for future investigation.

 The results presented here should be regarded as conservative, as other emulation schemes, such as \textit{sparse polynomial chaos expansion} \citep{Blatman_Sudret_2011,EuclidEmulator_etal_2018}, and further design optimisation \citep{Rogers_etal_2019, Caron_etal_2019} could help overcome the shortcomings of Gaussian process regression. Furthermore, we have only considered departures from \LCDM{} on scales of $k > 10^{-3} \, h/\rm{Mpc}$. This requirement could be relaxed by incorporating only one additional emulation parameter, $\Delta \bar{k}$, describing translations of the scale at which departures occur about this point. Since the scope of this preliminary study was to show the feasibility of pseudo cosmology emulation, these topics are also left to future work.

Our emulation method, coupled to the halo model reactions, enables future cosmic shear analyses by providing fast, accurate, and flexible predictions of the matter power spectrum deep in the non-linear regime, where we are more likely to find imprints of new physics beyond the concordance cosmology, if any \citep[see, e.g.,][]{Heymans_Zhao_2018}. Moreover, the approach developed in this work can be extended to predict other pseudo cosmology statistics, for instance galaxy clustering \citep{Zhai_etal_2019}, as well as mean halo properties, such as their abundance and concentration \citep{Kwan_etal_2013,McClintock_etal_2019}.

\section{Acknowledgements}

BG, MC and CH acknowledge support from the European Research Council under grant number 647112. BM acknowledges the support of a Principal's Career Development Scholarship from the University of Edinburgh. MC thanks Joey Faulkner for stimulating discussions in the early stages of this work.

\bibliographystyle{mnras}
\bibliography{references}

\appendix

\section{Gaussian process regression method} \label{Appendix:GPR}

The mathematics behind Gaussian process (GP) regression emulators have been covered extensively in previous work; we refer the interested reader to \cite{Rasmussen_Williams_2006} for a general discussion of GP, and to \cite{Habib_etal_2007} and \cite{Schneider_etal_2008} for its first applications to cosmology. Here we summarise its key points, which we implement using the publicly available code {\sc{scikit-learn}}\footnote{\url{https://scikit-learn.org}} \citep{scikit-learn}.

GP regression is a non-parametric Bayesian machine learning algorithm for constraining the distribution of functions consistent with observed data. Let us consider a data set, $\mathcal{D}$, consisting of $n$ realisations of an observable, $\boldsymbol{y}$, each corresponding to different values of the $d$-dimensional input parameters, $\boldsymbol{\pi} \in \mathbb{R}^d$, i.e. $\mathcal{D}=\{ (\boldsymbol{\pi}_i,y_i)|i=1,...,n \}$. The ensemble of matter power spectra measured from a suite of $N$-body simulations is a typical example of such a data set, and constitutes a training set for the emulator. The task of the GP emulator is to learn the distribution of functions, $f(\boldsymbol{\pi})$, which are consistent with the mapping between the training set input parameters - the `nodes' - and output, via

\begin{equation}
y(\boldsymbol{\pi}) = f(\boldsymbol{\pi}) + \epsilon_{\rm n}(\boldsymbol{\pi}) \,, 
\label{eqn:yModel}
\end{equation}

\noindent where $\epsilon_{\rm n}(\boldsymbol{\pi})$ is a noise term sampled from a mean-zero Gaussian distribution with a standard deviation given by the error on the training set observable, $y(\boldsymbol{\pi})$. The prediction, $y^*$, corresponding to an arbitrary coordinate $\boldsymbol{\pi^*}$, is then sampled from a generalisation of a Gaussian posterior probability distribution over the range of consistent functions\footnote{In this work we consider {\sc{halofit}} training predictions as a proxy for hypothetical numerical simulations with negligible error. Nevertheless, to prevent unwanted numerical instabilities we use the arbitrarily small constant default $\epsilon_{\rm n}$ in {\sc{scikit-learn}} for all $y(\boldsymbol{\pi})$.}. In other words, the GP emulator interpolates the observable from the input coordinates of the training set to the trial coordinates across a $d$-dimensional parameter space.

A key ingredient of our posterior is the Gaussian prior distribution of functions deemed to reasonably map between input and output. The prior is determined by a mean, conventionally taken to be zero, and a covariance function, known as the `kernel'. The kernel can take various functional forms, each described by a vector of hyperparameters, $\boldsymbol{h} = \{ A, p_1, \cdots, p_d\}$, governing the kernel's behaviour. Following \cite{Heitmann_etal_2009}, in this work we adopt the squared exponential form, which specifies the covariance between the functions $f(\boldsymbol{\pi})$ and $f(\boldsymbol{\pi^*})$ as

\begin{equation}
K(f,f^*; \boldsymbol{h}) \equiv  {\rm{cov}}\left( f(\boldsymbol{\pi}),f(\boldsymbol{\pi^*}) ; \boldsymbol{h} \right) = A \prod_{l=1}^d {\exp}\left[ \frac{(\pi_l - \pi^*_l) ^2}{p_l^2} \right] . \label{eqn:kernel}
\end{equation}

\noindent This kernel has the following properties: (1) the covariance varies smoothly within the parameter space; (2) it depends only on the Euclidean distance between points, such that $K(f,f^*; \boldsymbol{h}) = K(f^*,f; \boldsymbol{h})$; (3) predictions become maximally correlated when $\boldsymbol{\pi} = \boldsymbol{\pi^*}$; (4) the correlation is large for points in relative proximity and small for largely separated points; (5) each $p_l$ corresponds to the functions' characteristic length-scale of variation in each of the $d$ dimensions, whilst $A$ is the kernel amplitude.

The emulator is trained by finding values for the hyperparameters such that a distribution of functions which are optimally consistent with all realisations in the training set can be defined. In this work, we fit for these using the method built-in to {\sc{scikit-learn}}, which employs a gradient ascent optimisation of the marginal likelihood conditioned on the training set.

\section{Modelling BAO residuals} \label{Appendix:BAO_Modelling}

\begin{figure}
\begin{center}
\includegraphics[width=0.5\textwidth]{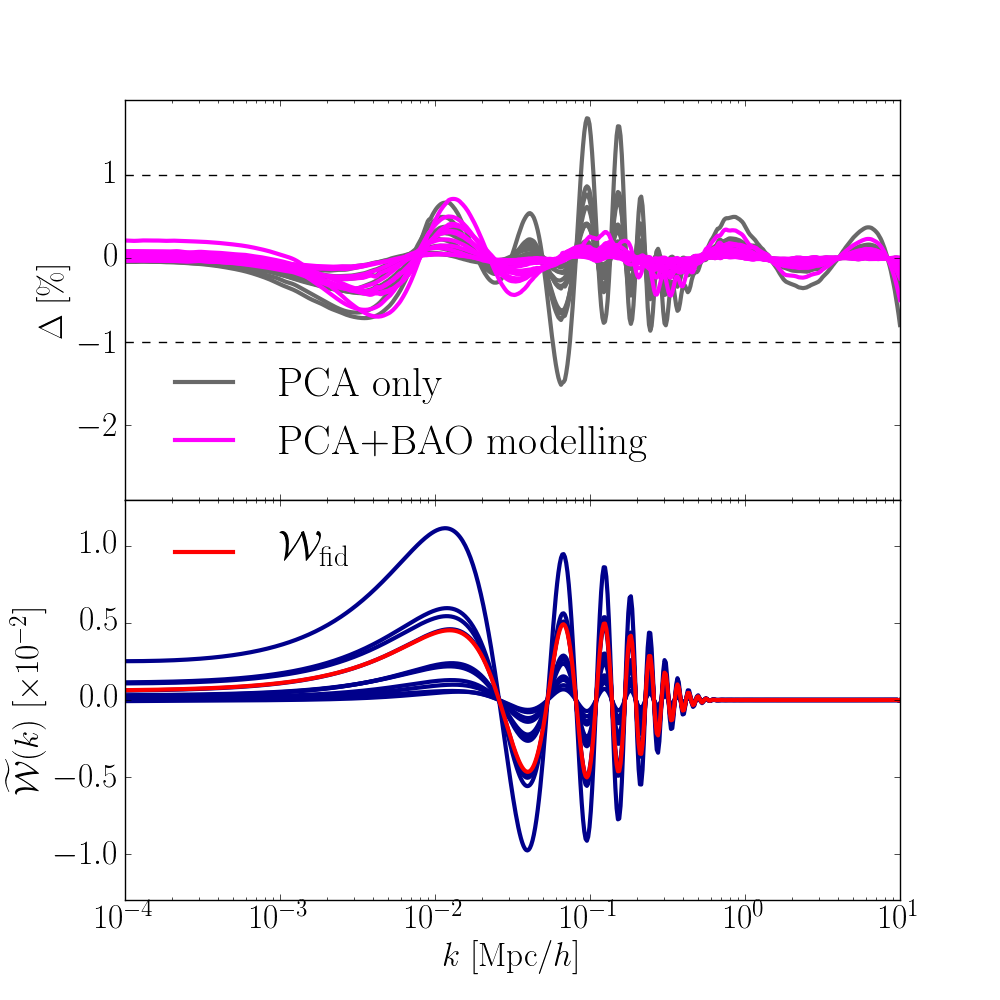} \\
\caption{\textit{Upper panel:} the accuracy of the shape reconstruction, $\Delta$, for the physical models featuring massive neutrinos (see Table \ref{tab:PhysMod_Table}). The grey curves show the results for direct PCA reconstruction using the 8 orthogonal basis functions derived in Section \ref{subsubsec:Random_Curves}. The magenta curves, reproduced from Figure \ref{fig:Rand_Curves}, illustrate the benefit of including the BAO residuals reconstruction via the two-step process in equation \ref{eqn:TwoStepShapeRecon}: the BAO residuals are first extracted with the de-wiggling algorithm presented in Appendix \ref{Appendix:Dewiggling} and then modelled using the template (see equation \ref{eqn:Wiggle_Mod}), whilst the PCA reconstruction is performed on the smooth component only. \textit{Lower panel:} the red curve shows the BAO residual associated with the MNU{\_}0.4 cosmology, which we use as the fiducial template. The blue curves show the best fit functions, $\widetilde{\mathcal{W}}(k; \bar{a}, \bar{b})$, to the oscillatory component of the shapes for the remaining physical models with massive neutrinos. All curves here correspond to $z=0$; results are very similar at $z=1$. } \label{fig:Dewiggling}
\end{center}
\end{figure}

As shown in Figure \ref{fig:PhysicalShapes}, models with massive neutrinos have shapes characterised by a rapidly oscillating component superimposed over a smooth function. These wiggles are associated with changes in the baryon acoustic oscillations (BAO) produced by differences in the sound horizon at the end of the drag epoch and shallower gravitational potentials before recombination~\citep{Hu_Sugiyama_1996,Hu_White_1996, Eisenstein_Hu_1998,Lattanzi_Gerbino_2017} compared to a massless neutrino \LCDM{} cosmology with the same standard parameters, i.e. $\boldsymbol{\pi}_\Lambda^{m_\nu=0} = \boldsymbol{\pi}_\Lambda^{m_\nu \neq 0}$, and $\Omega_{\rm c}^{m_\nu = 0} = \Omega_{\rm c}^{m_\nu \neq 0} + \Omega_{\nu}^{m_\nu \neq 0}$. Depending on the value of the standard cosmological parameters and neutrino mass, the amplitude of the BAO residual in the shape can be up to $\sim$1\% of the slowly varying component. The PCA reconstruction outlined in Section~\ref{subsec:GPR} requires a large number of components to simultaneously capture long- and short-scale variations. Hence we find errors $\gtrsim 1\%$ in reconstructing the $\sum m_\nu$-induced oscillations with only $n_\Phi=8$ basis functions, as shown by the grey curves in the upper panel of Figure \ref{fig:Dewiggling}.

In order to improve the reconstruction accuracy of the PCA, we isolate and remove the BAO residuals by applying the de-wiggling algorithm presented in Appendix \ref{Appendix:Dewiggling}. In short, the BAO in Fourier space maps to a localised bump in the discrete sine transform of the power spectrum. By removing this bump and performing the inverse transform, we obtain a smooth, ``de-wiggled" version of the power spectrum. This is performed on both the model featuring massive neutrinos and on the \LCDM{} baseline, before calculating the de-wiggled shape given by 
 
\begin{equation}
 S^{\rm{dw}}(k,z) = \frac{ P^{\rm{pseudo},\rm{dw}}_{\rm{L}}(k,z) }{ P_{\rm{L}}^{\Lambda,\rm{dw}}(k,z) } \,, \label{eqn:DewiggledShape}
\end{equation}

\noindent where the superscript `$\rm{dw}$' on the quantities on the right-hand side denote power spectra obtained from applying the de-wiggling algorithm. This smoothed component of the shape is captured very accurately by the basis functions. Reconstruction of the \textit{full} shape then rests on modelling the remaining oscillatory component, given by

\begin{equation} 
\mathcal{W}(k,z) = S(k,z) - S^{\rm{dw}}(k,z) . \label{eqn:WiggledShape}
\end{equation} 

\noindent Notice that all quantities in equations \ref{eqn:DewiggledShape} and \ref{eqn:WiggledShape} depend implicitly on the cosmological parameters. We experimented with fitting these oscillations with the function

\begin{equation}
\widetilde{\mathcal{W}}(k; a,b) = [a + b\log_{10}(k)]\mathcal{W}_{\rm fid}(k) \,, \label{eqn:Wiggle_Mod}
\end{equation} 

\noindent where $a$ and $b$ are free parameters controlling the amplitude and modulation of a fiducial wiggle template, $\mathcal{W}_{\rm fid}$, which we identify with the wiggle contribution of the MNU{\_}0.4 cosmology in Table \ref{tab:PhysMod_Table}, represented by the red curve in the lower panel of Figure \ref{fig:Dewiggling}. We checked that our fiducial template choice in equation \ref{eqn:Wiggle_Mod} is robust against changes in redshift, standard cosmological parameters, sum of neutrino masses and neutrino mass eigenstates. This ensures that the parameters $\{a,b\}$ are sufficient to capture BAO variations in the family of pseudo cosmologies treated in this work. The blue curves in the lower panel of Figure \ref{fig:Dewiggling} show the behaviour of our fitting formula, Eq.~\eqref{eqn:Wiggle_Mod}, for all of the test cosmologies in Table \ref{tab:PhysMod_Table} containing massive neutrinos. 

In summary, for each redshift and cosmology separately, we reconstructed the total shapes of the physical models as 
\begin{equation}
S(k; \boldsymbol{\pi}_\Lambda, \Delta\boldsymbol{\alpha}, a, b) = S^{\rm{dw}}(k;\boldsymbol{\pi}_\Lambda, \Delta\boldsymbol{\alpha}) + \widetilde{\mathcal{W}}(k; a,b) \,,
\label{eqn:TwoStepShapeRecon}
\end{equation}

\noindent where only the smooth, de-wiggled term is modelled via the PCA reconstruction given in equation \ref{eqn:Shape}. The magenta curves in the upper panel of Figure \ref{fig:Dewiggling} show the overall accuracy of the two-step reconstruction, originally presented in the lower panel of Figure \ref{fig:Rand_Curves}. We find that the inaccuracies in the BAO regime are greatly suppressed compared to performing the PCA directly on the full shape (grey curves). 

The fact that the cosmologies with massive neutrinos require two extra parameters to summarise their shapes to sub-per-cent accuracy, might imply that the emulator need be informed of $\{a,b\}$ to achieve this target accuracy in its predictions, thereby increasing the dimensionality from 13 to 15. To test this, we produced {\sc{MaxPro}} training sets optimised in 15 dimensions, with a range of wiggle parameters, $a$ and $b$, slightly broader than those obtained for the physical models with massive neutrinos. We then compared the emulator's performance with the physical models, when trained on these 15-dimensional sets, and when trained on the same sets but in 13-dimensions, omitting $a$ and $b$. We found, in fact, that the emulator is mostly insensitive to these two parameters, which therefore contribute only marginally to improving its accuracy. 

\begin{figure}
\begin{center}
\includegraphics[width=0.5\textwidth]{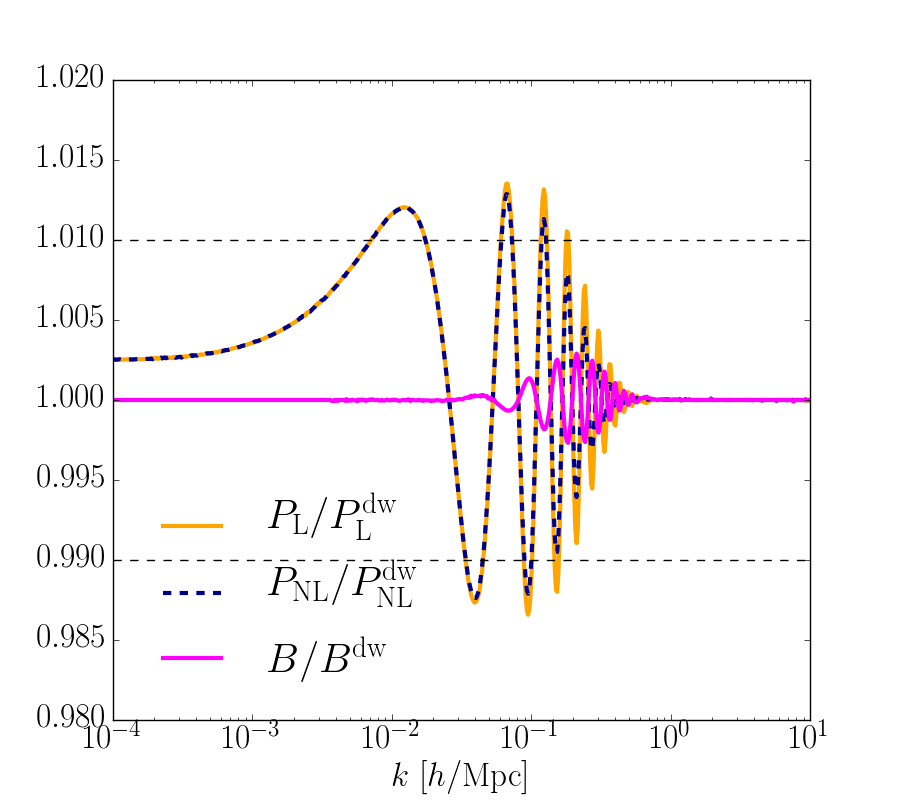} \\
\caption{Full to de-wiggled ratios for the boost factor, linear and non-linear matter power spectra of xMNU{\_}0.4 in Table \ref{tab:PhysMod_Table}. Quantities in the legend are defined as follows: $P_{\rm L} = S \times P_{\rm L}^\Lambda$; $P_{\rm L}^{\rm dw} = S^{\rm dw} \times P_{\rm L}^\Lambda$; $P_{\rm NL} = {\rm HF}[P_{\rm L}]; P_{\rm NL}^{\rm dw} = {\rm HF}[P_{\rm L}^{\rm dw}]$. The de-wiggled shape, $S^{\rm dw}$, is obtained as in equation \ref{eqn:DewiggledShape}, while HF is shorthand for {\sc halofit}. The solid orange and dashed blue lines represent the linear and non-linear BAO residuals, respectively, whose ratio is equivalent to the ratio of the boost factors, $B=P_{\rm NL}/P_{\rm L}$ and $B^{\rm dw}=P_{\rm NL}^{\rm dw}/P_{\rm L}^{\rm dw}$ (solid magenta). The contribution of the BAO residual to the non-linear power spectrum exceeds 1\%, but is at most $\simeq$0.3\% in the boost, which implies we can safely ignore the wiggle parameters, $\{a,b\}$, in emulation.} \label{fig:Ratio_Wigs_To_DeWigs}
\end{center}
\end{figure}

Figure \ref{fig:Ratio_Wigs_To_DeWigs} holds the key to this somewhat surprising result, where we adopt the xMNU{\_}0.4 cosmology as a worst case scenario among our physical  models. Plotted in solid orange is the $S/S^{\rm dw}$ ratio, which is equivalent to the ratio of pseudo linear power spectra with or without changes to the BAO scale compared to \LCDM{} (see equation \ref{eqn:DewiggledShape}). In dashed dark blue is the ratio of their non-linear counterparts obtained with {\sc halofit}. Solid magenta shows the ratio between the boost factors derived from using the full or de-wiggled shape, which is also equal to the ratio of the dark blue and orange curves. We see that although the BAO residual amounts to $\sim$1\% of the non-linear power spectrum, its contribution to the boost factor is much smaller. In turn, this means that the emulator can reconstruct the boost factor with the required accuracy irrespective of whether the information on the BAO residual is provided or not. In other words, the boost factor effectively damps down the correction associated with changes to the BAO caused by massive neutrinos, eliminating the need for the additional wiggle modelling (equation \ref{eqn:Wiggle_Mod})\footnote{Although we verified this statement with {\sc halofit} we expect it to remain valid in simulations as well, assuming that {\sc halofit} provides accurate predictions for the ratio $P_{\rm NL}/P_{\rm NL}^{\rm dw}$.}.

The fact that early-time modifications to the BAO physics have such a weak effect on the boost factor is a strength of our emulation scheme; an emulator designed to directly predict the pseudo non-linear power spectra is likely to incur $\gtrsim 1\%$ errors unless detailed modelling of the BAO residual is employed. The unimportance of the wiggle parameters means that the 13-dimensional emulator is sufficient to achieve our target accuracy. In Appendix \ref{Appendix:Neff}, we investigate the performance of the emulator with cosmologies featuring extra relativistic degrees of freedom, for which changes in the acoustic oscillations prior to the epoch of last scattering leave especially prominent wiggles in the late-time shape ratios. We find the emulator achieves sub-per-cent accuracy at $z=0$, but less consistent results at higher redshift for this model.

Although modelling of the BAO residuals proves to be unnecessary, their extraction through the de-wiggling algorithm remains important. By acting as a low-pass filter, the PCA reconstruction can in principle remove rapid oscillations and capture the broadband shape. However, the PC weights thus derived are somewhat different from those obtained after de-wiggling, generating an error that ultimately propagates to the boost factor. We compare the emulator performance with and without de-wiggling in Appendix \ref{Appendix:Emu_Without_DW}.

\section{De-wiggling methodology} \label{Appendix:Dewiggling}

\begin{figure}
\begin{center}
\includegraphics[width=0.5\textwidth]{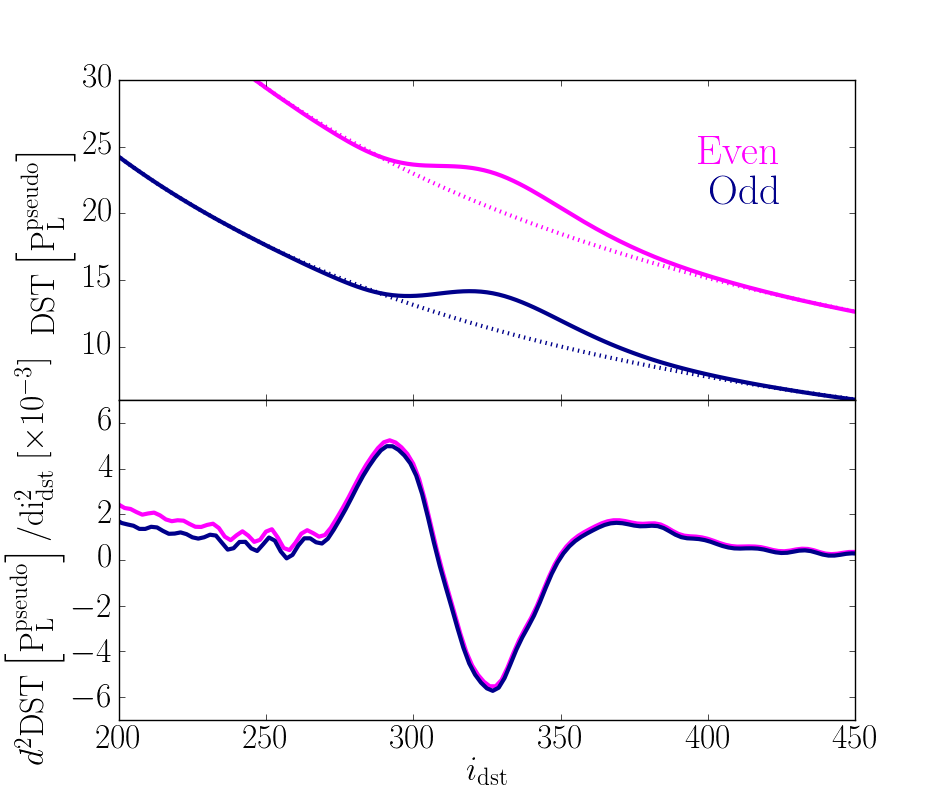} \\
\caption{\textit{Upper panel:} The discrete sine transform (DST) of the pseudo linear matter power spectrum for the F4-MNU{\_}0.4 model, split into two arrays with even (magenta) and odd (dark blue) indices, $i_{\rm dst}$, before (solid lines) and after (dashed lines) the bump is removed. The bump is a localised manifestation of the BAO wiggles present in the power spectrum. \textit{Lower panel:} The second derivative of the DST used to identify the $i_{\rm dst}^{\rm{min}}$ and $i_{\rm dst}^{\rm{max}}$ values which bracket the bump in the DST. We adjust the values of these for each model, optimising the smoothness of the returned de-wiggled shape. The final $i_{\rm dst}^{\rm{min}}$ and $i_{\rm dst}^{\rm{max}}$ values are close to, respectively, the local minimum on the left-hand side of the centre of the prominent oscillation in the second-derivative, and the local maximum on the right-hand side. } \label{fig:DST_DW}
\end{center}
\end{figure}

Here we outline in more detail the method, discussed briefly in Appendix \ref{Appendix:BAO_Modelling}, for de-wiggling linear matter power spectra following \citet{Baumann_etal_2018}. This procedure is performed on the pseudo and $\Lambda$CDM spectra separately, and therefore we denote both with $P_{\rm{L}}(k)$ in the following: 

\begin{enumerate}
\item We firstly interpolate the linear matter power spectra featuring BAO wiggles, $P_{\rm{L}}(k)$, from the logarithmically-spaced $k$ values onto a linear-spaced array, $k_n$, sampled at $2^n$ points, where $n$ is an integer. This is so as to improve the computational efficiency of the Fast Discrete Sine Transform (DST), which is then performed on $\log_{10}[k_nP_{\rm{L}}(k_n)]$ using the orthonormalised type-II sine transform. The resulting array, which has indices denoted by $i_{\rm dst}$ and length $2^n$, is split into two separate arrays, one with the even $i_{\rm dst}$ indices and the other with odd $i_{\rm dst}$ indices.  

\item The DST of the even and odd arrays each feature a bump, which is a localised manifestation of the BAO in the $k$-space of the matter power spectrum. This is shown by the solid lines for one of the physical models in the upper panel of Figure \ref{fig:DST_DW}. In order to de-wiggle the power spectrum, we must remove these bumps. To this end, we take the second derivative of the DST, converting the observed bump in each array into a prominent single-period oscillation, as shown by the lower panel of Figure \ref{fig:DST_DW}. We then identify $i_{\rm dst}^{\rm{min}}$ and $i_{\rm dst}^{\rm{max}}$ values bracketing the oscillation, which are close to the local minimum on the left-hand side of the approximate centre of this feature and the local maximum on the right-hand side respectively.

\item We remove the corresponding $i_{\rm dst}$-range from the even and odd DST arrays, multiply each by $(i_{\rm dst}+1)^2$ and interpolate across the gap between $i_{\rm dst}^{\rm{min}}$ and $i_{\rm dst}^{\rm{max}}$ with cubic splines. The $(i_{\rm dst}+1)^2$ factors are then divided out, generating smooth ``de-bumped" even and odd DST arrays, where the information corresponding to the BAO has been removed. This is shown by the dashed lines in the upper panel of Figure \ref{fig:DST_DW}.    

\item We recombine the even and odd DST arrays into a single array. Performing an inverse type-II DST produces the logarithmic de-wiggled linear matter power spectra weighted by the linear-spaced $k_n$ array, $\log_{10}[k_nP^{\rm{dw}}_{\rm{L}}(k_n)]$. From this we obtain the de-wiggled linear spectra itself, $P^{\rm{dw}}_{\rm{L}}(k_n)$.  

\item After performing this on both the pseudo and $\Lambda$CDM power spectra, we take the ratio of the respective $P^{\rm{dw}}_{\rm{L}}(k_n)$ to obtain $S^{\rm{dw}}(k_n)$, the de-wiggled shape sampled at the linearly-spaced wavenumbers. Finally we interpolate the de-wiggled shape back to the logarithmically-spaced $k$ sampling, obtaining $S^{\rm{dw}}(k)$. We find that there is less numerical noise at the upper and lower wavenumber bounds of the de-wiggled shape when we take the ratio of the linearly-sampled de-wiggled power spectra before interpolating, rather than the other way around. 

\end{enumerate}

Whereas \cite{Baumann_etal_2018} chose fixed values of the $i_{\rm dst}^{\rm{min}}$ and $i_{\rm dst}^{\rm{max}}$ relative to the oscillation in the second derivative of the DST arrays, we optimise for these parameters in our analysis with the physical models. Specifically, we cycle through a grid of $(i_{\rm dst}^{\rm{min}}, i_{\rm dst}^{\rm{max}})$ combinations which are centred on the local minimum and maximum on either side of the oscillation, and identify those which return the smoothest de-wiggled shape. The smoothness is measured by integrating the second derivative of $S^{\rm{dw}}(k)$, convolved with a Gaussian filter, in the range of the BAO wiggles. The optimal $(i_{\rm dst}^{\rm{min}}, i_{\rm dst}^{\rm{max}})$ combination is that which minimises this integral. We find the de-wiggled shapes are insensitive to the width of the Gaussian filter, which is employed so as to ensure numerical noise does not bias identification of the optimal $(i_{\rm dst}^{\rm{min}}, i_{\rm dst}^{\rm{max}})$ range.

\section{Extension to extra relativistic degrees of freedom} \label{Appendix:Neff}

Appendix \ref{Appendix:BAO_Modelling} demonstrated the unimportance of modelling the BAO residuals in the shapes of the physical cosmologies, justifying the 13-dimensional emulation scheme. In order to further verify the validity of this, we test our emulator scheme against a theoretically motivated cosmology with an especially prominent BAO residuals in the shape ratio. 

\begin{figure}
\begin{center}
\includegraphics[width=0.5\textwidth]{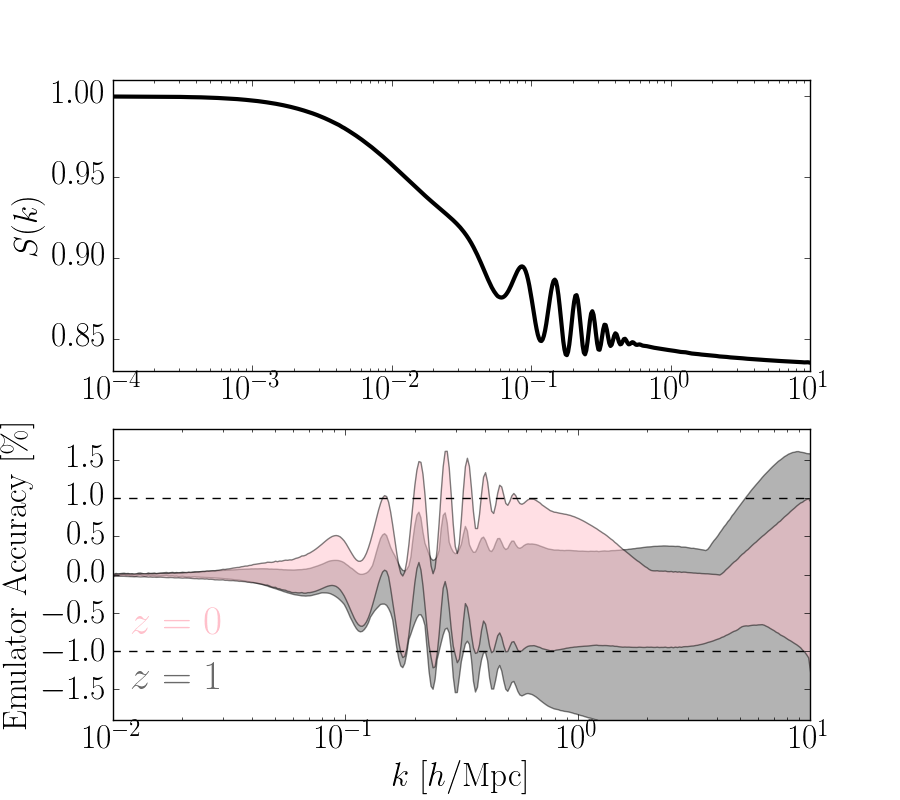} \\
\caption{\textit{Upper:} the $z=0$ shape ratio for the model with $\Delta N_{\rm eff}=1$ extra relativistic degrees of freedom. The BAO residual is at the level of $\simeq$2\%, about twice as large as the residuals for physical models with massive neutrinos, shown in Figure \ref{fig:PhysicalShapes}. \textit{Lower:} the full range of emulation accuracies achieved with the 10 realisations of the $N=500$ training set at $z=0$ (pink) and $z=1$ (grey). Despite the larger prominence of the BAO residual to the shape, we find its contribution to the boost factor remains at the sub-per-cent level, facilitating reconstruction of the $z=0$ pseudo non-linear power spectrum to better than the target accuracy with almost all of the training set realisations. The larger inaccuracies observed at $z=1$ suggest that this model cannot reliably be constructed at higher redshifts with the current emulation scheme.} \label{fig:Neff} 
\end{center}
\end{figure}

The existence of a relic sea of neutrinos with effective relativistic degrees of freedom, $N_{\rm eff}=3.046$, is a general prediction of the standard model \citep{Mangano_etal_2005}. However, physics beyond this paradigm includes scenarios with additional relativistic particles (or `dark radiation') at the epoch of decoupling, typically quantified by the variation $\Delta N_{\rm eff}$, such that  
\begin{equation}
N_{\rm eff} = 3.046 + \Delta N_{\rm eff} .
\end{equation}
This has the effect of reducing the expansion rate and increasing the acoustic scale \citep{Archidiacono_etal_2013}, thereby acting to ease the tension between the early- and late-time measurements of the Hubble constant \citep{Riess_etal_2019}. We consider a model with $\Delta N_{\rm eff} = 1$ extra relativistic degrees of freedom, which lies several standard deviations away from the best-fit result of the \cite{Planck_2018} analysis. The BAO residuals in the shape of this cosmology at $z=0$ are shown in the upper panel of Figure \ref{fig:Neff}.

Following the same procedure employed for the physical models in Table \ref{tab:PhysMod_Table}, we obtain a smoothed version of this shape using the de-wiggling algorithm (see Appendix \ref{Appendix:Dewiggling}), perform a PCA decomposition to obtain the PCA weights, $\Delta\boldsymbol{\alpha}$, but refrain from modelling the BAO residual component as described in Appendix \ref{Appendix:BAO_Modelling}. We then infer the $P^{\rm pseudo}_{\rm NL}$ for this model using the emulator in the fiducial 13-dimensional setup and compare to the prediction from {\sc halofit}.

The results using the 500-node training sets are shown in the lower panel of Figure \ref{fig:Neff}. We find that despite the BAO residual in the shape being as large as $2\%$ for this cosmology\footnote{For comparison, this feature never exceeds 1\% in our other test cosmologies.}, the emulator reproduces the $P^{\rm pseudo}_{\rm NL}$ to better than 1\% accuracy at $z=0$ (pink) in almost all of the training set realisations. The reason that these accuracies are achievable, as with the physical models, is due to the boost factor suppressing the early-time modifications of the BAO to sub-per-cent levels. We also note that changes in the linear matter power spectrum for models with $\Delta N_{\rm eff} \neq 0$ (see Figure \ref{fig:Neff}) are caused by processes happening well before the matter dominated epoch. At redshifts $\gtrsim 100$ the background and growth evolution in these extensions are indistinguishable from those in \LCDM{}. Therefore, the \textit{full} non-linear matter power is entirely captured by the pseudo cosmology (i.e. the reaction is unity for all scales, see Section \ref{subsec:C19}). At $z=1$, however, we find that this model cannot be reliably predicted to within the target accuracy across the training set realisations. This is perhaps an indication that models with such marked BAO residuals are beyond the capabilities of our emulator as specified in this work. This may be improved upon given an alternative emulation method, such as \textit{sparse polynomial chaos expansion} \citep{Blatman_Sudret_2011, EuclidEmulator_etal_2018}, or with careful optimisation of the training set configuration \citep{Rogers_etal_2019, Caron_etal_2019}.

\section{Emulation without de-wiggling} \label{Appendix:Emu_Without_DW}

\begin{figure}
\begin{center}
\includegraphics[width=0.5\textwidth]{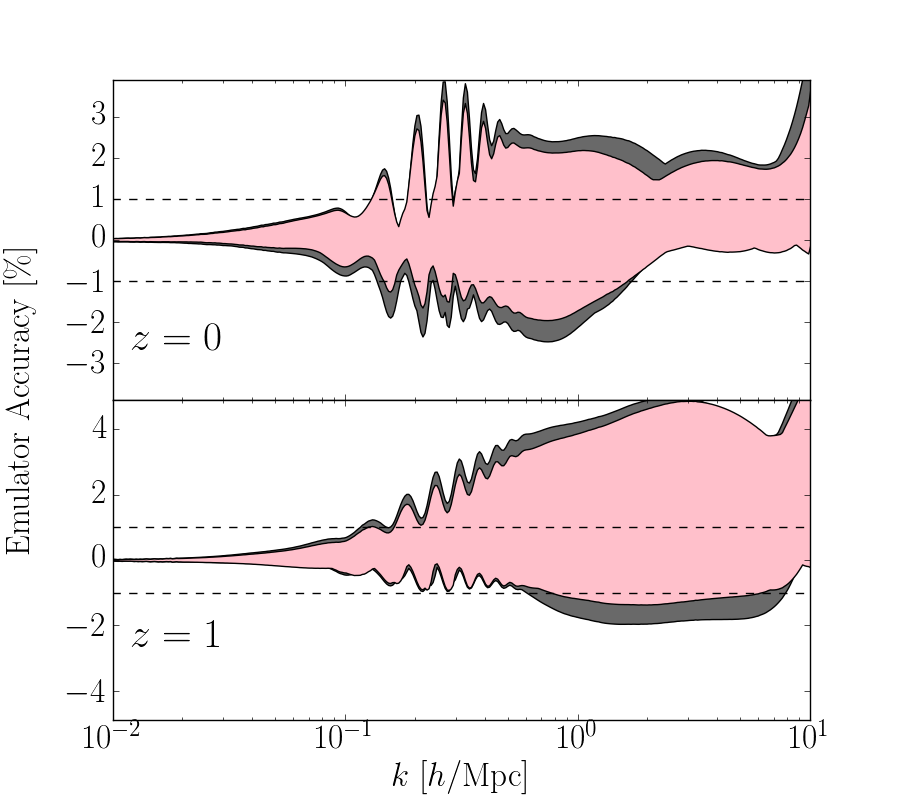} \\
\caption{The range of accuracies of the  emulated pseudo matter power spectra for the xMNU{\_}0.4 cosmology (see Table \ref{tab:PhysMod_Table}) at $z=0$ (upper panel) and $z=1$ (lower panel), obtained with our 10 training sets of 500 nodes. The inner (pink) band corresponds to our fiducial emulation method, reproduced from Figure \ref{fig:Model_Accuracy}. The outer (grey) band corresponds to the case with no de-wiggling and PCA decomposition performed directly on the full shape.} \label{fig:Emu_No_Dewiggle} 
\end{center}
\end{figure}

Although modelling of the BAO residuals in the shapes to obtain the wiggle parameters, $a$ and $b$, has negligible effect on the results (see Appendix \ref{Appendix:BAO_Modelling}), we find that shape de-wiggling remains important for improving emulation accuracy. This is because the PCA is a low-pass filter, effective at capturing only the low-frequency component of the shapes. This property is imparted by the basis set, $\Phi_i$, derived from smooth curves (see Section \ref{subsubsec:Random_Curves}), such that the corresponding weights, $\Delta \alpha_i$, are designed to contain smooth component information. Consequently, performing the PCA directly on the full shape leads to misestimation of the $\Delta \alpha_i$ values, and bias in the emulation.

Figure \ref{fig:Emu_No_Dewiggle} presents the effect of this for the xMNU{\_}0.4 cosmology (see Table \ref{tab:PhysMod_Table}) at $z=0$ (upper panel) and $z=1$ (lower panel). The pink band shows the full range of emulation accuracies achieved with the 10 realisations of the 500-node training set for our fiducial emulation method incorporating the de-wiggling, reproduced from Figure \ref{fig:Model_Accuracy}. The grey band shows this quantity for the case with no de-wiggling, where the PCA decomposition is performed directly on the full shape. Indeed, excess inaccuracies of $\sim$0.5\% are observed in the absence of de-wiggling, for the offset PCA weights map to a boost factor related to a slightly different shape compared to the {\sc halofit} predictions. Although currently subdominant, these errors will become more relevant for future emulators designed to have target performance of $\sim$0.5\%, a requirement imposed by irreducible inaccuracies in the halo model reactions.
 
\newpage

\end{document}